\documentclass{aa} 
\usepackage{hyperref}
\usepackage{graphicx}
\usepackage{txfonts}
\usepackage[separate-uncertainty=true, multi-part-units=single, print-unity-mantissa=false]{siunitx}
\DeclareSIUnit\erg{erg}

\def\MgII{\ion{Mg}{II}}
\def\OIII{[\ion{O}{III}]}
\def\OII{[\ion{O}{II}]}
\def\CIV{\ion{C}{IV}}
\def\SiIV{\ion{Si}{IV}}
\def\CIII{\ion{C}{III}]}
\def\CaII{\ion{Ca}{II}}
\def\CI{\ion{C}{I}}

\begin{document}

\title{On the dusty proximate damped Lyman-$\alpha$ system toward Q\,2310$-$3358 at $z=2.40$}\titlerunning{}

\author{
S.~Han\inst{1,2},
J.-K.~Krogager\inst{3,4},
C.~Ledoux\inst{5},
G.~Ma\inst{1,2},
K.~E.~Heintz\inst{1,2},
S.~J.~Geier\inst{6,7},
L.~Christensen\inst{1,2},
P.~M\o ller\inst{1},
J.~P.~U.~Fynbo\inst{1,2}
}
\institute{
Cosmic DAWN Center,
\email{jfynbo@nbi.ku.dk}
\and
Niels Bohr Institute, University of Copenhagen, Jagtvej 155, 2200 Copenhagen N, Denmark
\and
French-Chilean Laboratory for Astronomy, IRL 3386, CNRS and U. de Chile, Casilla 36-D, Santiago, Chile
\and
Centre de Recherche Astrophysique de Lyon, Universit{\'e} de Lyon 1, UMR5574, 69230 Saint-Genis-Laval, France
\and
European Southern Observatory, Alonso de C\'ordova 3107, Vitacura, Casilla 19001, Santiago, Chile
\and
Instituto de Astrof{\'i}sica de Canarias, V{\'i}a L{\'a}ctea, s/n, 38205, La Laguna, Tenerife, Spain
\and
Gran Telescopio Canaias (GRANTECAN), 38205 San Crist{\'o}bal de La Laguna, Tenerife, Spain
}
\authorrunning{Han et al.}

\date{Received 2025; accepted, 2025}

\abstract{Quasar absorption systems not only affect the way quasars are selected, but also serve as key probes of galaxies, providing insight into their chemical evolution and interstellar medium (ISM). Recently, a method based on Gaia astrometric measurements has aided the selection of quasars reddened by dust hitherto overlooked. We conducted a spectroscopic study using VLT/X-Shooter on one such dust-reddened quasar, Q\,2310-3358. This quasar, at $z = 2.3909\pm0.0022$, is associated with a damped Lyman-$\alpha$ absorber (DLA) at nearly the same redshift $2.4007\pm0.0003$, with a neutral hydrogen column density of $\log N(\ion{H}{I})=21.214\pm0.003$. The DLA is very metal-rich (close to solar metallicity after correction for depletion on dust grains). Its properties align with the metal-to-dust ratio and the mass-metallicity relation established in previous large samples of DLAs. Surprisingly, given its proximity to the quasar in redshift, the absorber has strong cold gas characteristics, including {\CI} and H$_2$. Based on the derived kinetic temperature of $71^{+28}_{-15}$~K, we infer the presence of a strong UV radiation field, which in turn suggests that the quasar and the DLA are in close proximity, i.e., part of the same galaxy and not just different objects in the same overdensity of galaxies. We used the line ratios of the {\CI} fine-structure lines to constrain the density of the cold gas, yielding $n_{\rm H} \sim 10^{3}~\mathrm{cm}^{-3}$. Our analysis extends the understanding of $z_{abs} \approx z_{em}$ absorption line systems and provides valuable constraints on the interplay between dust, metals, and neutral gas in the ISM of early galaxies.

}    


\maketitle

\section{Introduction} 

\label{sec:introduction}
Ever since their discovery, quasi-stellar objects (quasars) have served as key probes for investigating a range of important aspects of the distant universe \citep{1963Natur.197.1040S,1964ApJ...140....1G}. The selection of complete samples of quasars from survey data, therefore, has long been a critical issue to avoid biases. Photometric selection criteria, due to their efficiency and relative accuracy, have been widely used and continuously improved \citep{1965ApJ...142.1307S,2000MNRAS.312..827W,2012ApJ...753...30S}.

In particular, dust-obscured quasars are both fainter and reddened, resulting in a dust bias that makes them more difficult to identify through photometric selection than unobscured quasars \citep{1989ApJ...337....7F,1991ApJ...378....6P,2009MNRAS.393..557P,2019MNRAS.486.4377K}. The dust causing the extinction is located in a range of different locations, either close to the central black hole, further out in the interstellar medium (ISM) of the host galaxy, in intervening galaxies, or in the Milky Way disk. Dust in the host galaxy or in intervening galaxies will typically be associated with strong hydrogen absorption line systems. When the neutral hydrogen column density of the absorber is sufficiently high (higher
than $10^{20.3}$ cm$^{-2}$), these absorbers are called damped Ly$\alpha$ systems (DLAs). Damped Ly$\alpha$ systems can contain substantial amounts of dust and exhibit high metal abundance, therefore playing a crucial role in the study of chemical evolution and the ISM of galaxies \citep{1986ApJS...61..249W,1990AuJPh..43..227P}. New selection methods have been proposed, with the aim of finding quasars missed by more biased selection methods, with several successful advances reported \citep[e.g., ][]{1995Natur.375..469W,Glikman2013,2016ApJ...832...49K,2016MNRAS.455.2698K,2018A&A...615A..43H, Glikman2018, Glikman2022}.

In addition to photometry, an innovative method for selecting more complete samples of quasars has been proposed. This approach combines Gaia astrometric measurements with photometric selection criteria in the optical and in the near-infrared, facilitating the construction of a more complete quasar sample \citep{2015A&A...578A..91H,2018A&A...615L...8H,2020A&A...644A..17H,2019A&A...625L...9G}. Using this approach, we selected a quasar associated with an absorber at the same redshift. Such systems, so-called proximate systems, are relatively rare among all DLAs \citep{Weymann1977, Moller1998, Ellison2010}. In cases where the DLA lies in close proximity to the quasar, the detection of molecules has been considered of particular interest \citep{1998A&A...335...33S,Noterdaeme2019, 2020MNRAS.497.1946B,2025Natur.641.1137B}.

In this paper, we study a dust-obscured quasar, in which a DLA has been detected at nearly the same redshift as the quasar. We observed this DLA using X-Shooter on the Very Large Telescope (VLT) and performed a spectroscopic analysis. In Sect.~\ref{sec:data}, a new spectroscopic observation of Q\,2310$-$3358 at $z = 2.40$ is presented, showing  prominent features associated with a Lyman-$\alpha$ absorber. In Sect.~\ref{sec:results}, we show the results of this DLA, including absorption lines, extinction, metallicity, and molecules. In Sect.~\ref{sec:discussion and conclusions}, we present our discussion and conclusions.

\section{Observations and data reduction}    \label{sec:data}

The quasar Q\,2310--3358, located at right ascension (R.A.) 23:10:15.2 and declination (Dec.) $-33$:58:10.2 (J2000), was identified using a standard photometric selection criterion in combination with a novel Gaia-based astrometric method. In the quasar catalog, we used a selection criteria similar to that of \citet{2019A&A...625L...9G}. First, in terms of astrometry, we considered total proper motions $\mu$ consistent with zero within $2\sigma$ \citep{2018A&A...615L...8H}. In terms of photometry, we selected sources with $u - g > 1$ and $r - z > 0.5$ \citep{2018A&A...615A..43H}. The photometric data for Q\,2310--3358 were obtained from several large-scale surveys, including the Kilo Degree Survey \citep[KiDS;][]{2013ExA....35...25D}, the VISTA Kilo-degree Infrared Galaxy survey \citep[VIKING;][]{2013Msngr.154...32E}, and the Wide-field Infrared Survey Explorer \citep[WISE;][]{2010AJ....140.1868W}. These datasets provide broad wavelength coverage from the optical to the near-infrared. The magnitudes in each band are listed in Table~\ref{tab:phot}. This source exhibits colors of $u-g = 2.151$, $g-r = 0.744$, $r-z = 0.35$, and $J-K = 1.50$.

\begin{table}[!t]
\centering
\begin{minipage}{0.5\textwidth}
\centering
\caption{Photometry of Q\,2310--3358.}
\begin{tabular}{lll}
\noalign{\smallskip} \hline \hline \noalign{\smallskip}
Band  & AB Magnitude & Error \\
\hline
$u$ & 22.70 & 0.12  \\
$g$ & 20.651 & 0.007 \\
$r$ & 19.907 & 0.004  \\
$Z$ & 19.56 & 0.01   \\
$Y$ & 18.18 & 0.02   \\
$J$ & 17.80 & 0.03   \\
$H$ & 17.35 & 0.04   \\
$K_{\mathrm{s}}$ & 16.30 & 0.03   \\
$W1$ & 15.69 & 0.06   \\
$W2$ & 14.76 & 0.08  \\
$W3$ & 11.14 & 0.01   \\
$W4$ & 8.97 & 0.47   \\
\hline
\noalign{\smallskip} \hline \noalign{\smallskip}
\end{tabular}
\centering
\label{tab:phot}
\end{minipage}
\end{table}

We here present observations of Q\,2310--3358 obtained with the medium-resolution spectrograph X-shooter on board VLT. The quasar was observed at three different position angles (0, 60$^\mathrm{o}$, and $-60^\mathrm{o}$ East of North). According to \citet{2010MNRAS.408.2128F}, with this strategy, 90\% of the galaxy counterparts of intervening DLAs at redshifts around 2--3 can be covered by at least one slit, and observations with all slits also contribute to a good spectrum of the quasar.

\begin{figure*}[th]
    \centering
    \includegraphics[scale=0.40]{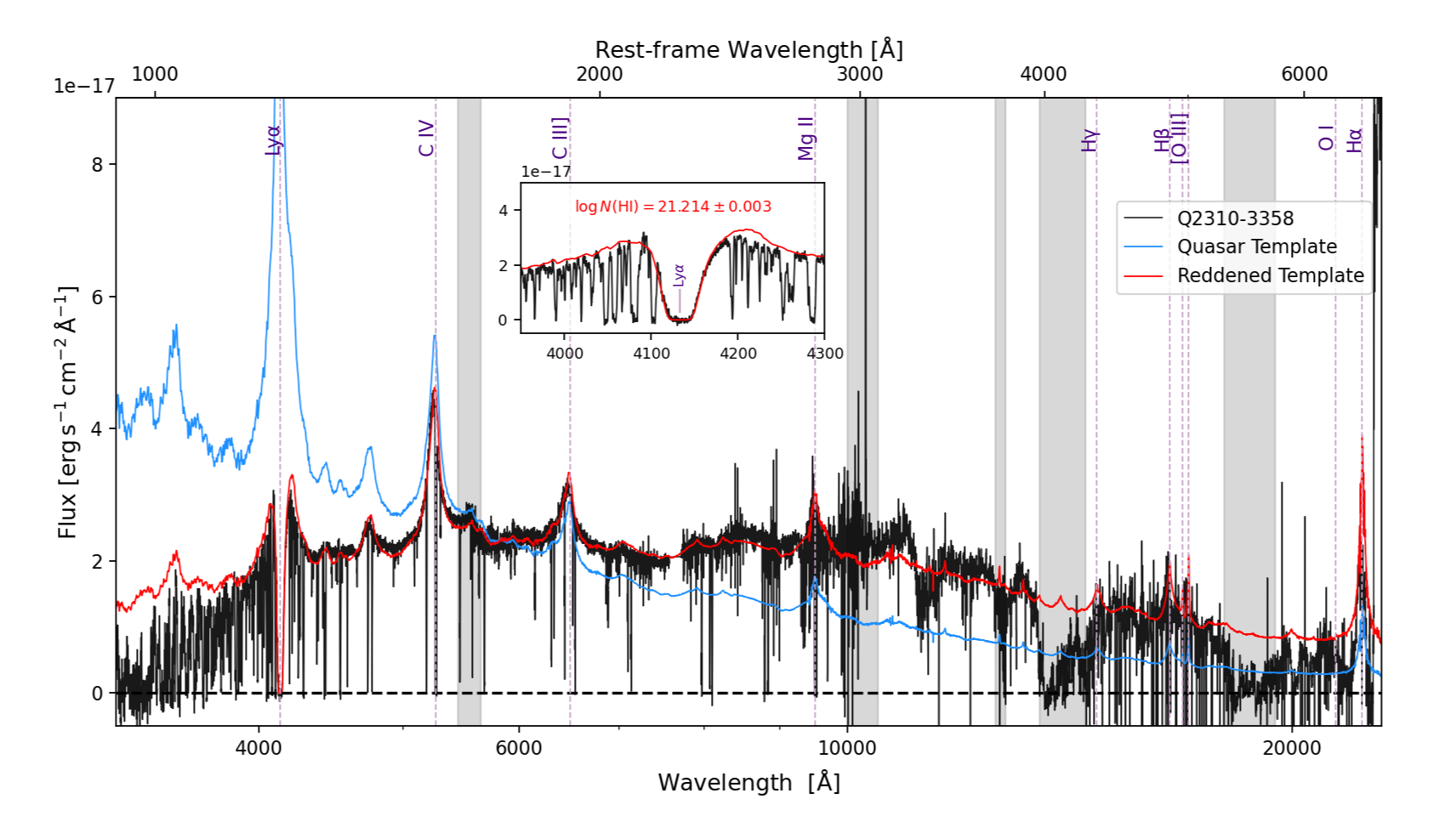}
    \caption{Spectrum of Q\,2310-3358 obtained with VLT/X-Shooter (black curve), with the positions of the prominent quasar emission lines indicated by dashed purple lines. The blue curve represents the composite quasar spectrum from \citet{2016A&A...585A..87S}. A reddened template is plotted in red using the best-fit extinction parameters, along with the addition of a Ly\(\alpha\) absorption line corresponding to $\log N(\ion{H}{I}) = 21.21\pm0.003$. The inset provides a zoomed-in view of the Ly\(\alpha\) line from the absorber. Gray bands mark the regions where the spectrum is unreliable due to X-Shooter arm overlaps and telluric absorption.}
    \label{Spectrum}
\end{figure*}

X-shooter provides continuous spectral coverage from near-ultraviolet (300 nm) to near-infrared (2500 nm). Both 1D and 2D spectra are available from the standard ESO pipeline reduction of the raw data \citep{2010SPIE.7737E..28M}. For each arm of the data, the spectra from all three position angles were combined using inverse-variance weighting to produce the combined spectrum shown in Fig. \ref{Spectrum}. The instrument acts as an atmospheric dispersion corrector, meaning that this strategy can also be used well away from the zenith. The log of observations can be seen in Table \ref{tab:log}.

\begin{table}[!t]
\centering
\begin{minipage}{0.5\textwidth}
\centering
\caption{Log of observations.}
\begin{tabular}{llrll }
\noalign{\smallskip} \hline \hline \noalign{\smallskip}
Date  & Exp. time & PA & Seeing & Airmass \\
      &   (sec)       & (degree)  &  (arcsec)  &  \\
\hline
08/06/2024 & 3300 & 0   & 1.18 & 1.05 \\
15/06/2024 & 3300 & $-60$  & 1.69 & 1.03 \\
16/06/2024 & 3300 & 0   & 1.25 & 1.14 \\
25/06/2024 & 3300 & 60 & 1.25 & 1.05 \\
\hline
\noalign{\smallskip} \hline \noalign{\smallskip}
\end{tabular}
\centering
\label{tab:log}
\end{minipage}
\end{table}

\section{Results}    \label{sec:results}

\subsection{Emission and absorption lines}
We first aim to determine the systemic redshift of the quasar. As shown in Fig.~\ref{Spectrum}, we can easily identify the {\SiIV}, {\CIV}, and {\CIII} emission lines in the X-shooter spectrum. However, these high ionization broad emission lines are known to be blueshifted \citep{1992ApJS...79....1T,2016ApJ...831....7S}, which means that these lines underestimate the systemic redshift of the host galaxy. Given that we cannot clearly see any of the narrow emission lines, such as the {\OII} and {\OIII} lines, nor the strong stellar absorption line {\CaII}\,$\lambda$3934 (K), which are the most accurate redshift tracers according to \citet{2010MNRAS.405.2302H} and \citet{2016ApJ...831....7S}, we need to rely on the low ionization broad emission lines.

\begin{figure}[th]
\centering
\includegraphics[scale=0.45]{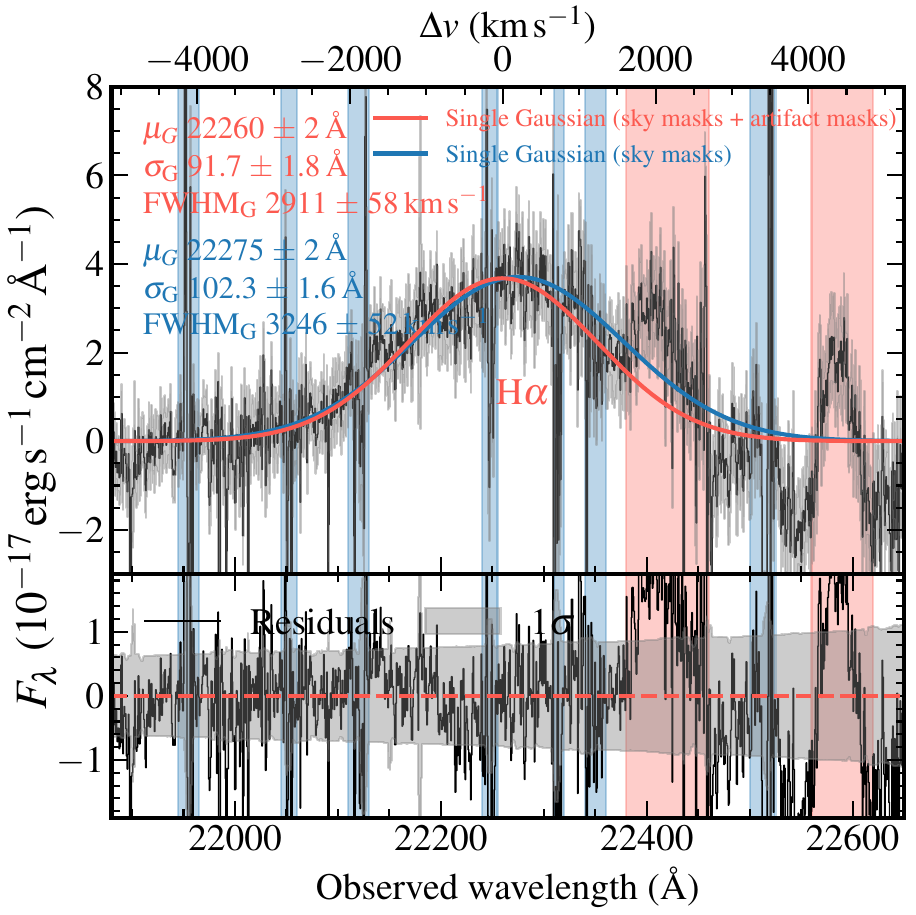}
\caption{Upper panel: Two single-Gaussian profiles fit to the H$\alpha$ emission line. The red curve shows the fit after masking the artifacts (marked by the red bands), 
while the blue curve displays the result when the artifacts are not masked. Bottom: Fitting residuals, where the gray shading denotes the $1\sigma$ uncertainty of the spectrum. In both cases, the sky-line regions are masked out, as indicated by the blue bands. The mean ($\mu_G$), standard deviation ($\sigma_G$), and $\mathrm{FWHM}_G$ (corrected for the instrument resolution of \SI{39.377}{\km\per\s}) of the fit are listed in the plot.}
\label{fig:H_alpha}
\end{figure}

Redshifts determined from low ionization broad emission lines are also found to be quite close to the systemic redshift \citep{2010MNRAS.405.2302H,2016ApJ...831....7S}. {\MgII}, ${\mathrm{H} \beta}$, and {\OIII} lines are heavily embedded in the telluric lines and sky subtraction residuals in our spectra, while the ${\mathrm{H} \alpha}$ line is detected at the end of the spectrum in the K band. We determined the systemic redshift of this quasar based on this ${\mathrm{H} \alpha}$ line. A linear continuum was fit only based on the spectrum at wavelengths shorter than the ${\mathrm{H} \alpha}$ line and subtracted from the spectrum. Though a combination of several Gaussian functions are often applied to fit the emission lines, such as in \cite{2016ApJ...831....7S}, we found that one single Gaussian fit reproduced the ${\mathrm{H} \alpha}$ line profile quite well.
The fitting result is shown in Fig.~\ref{fig:H_alpha}. To account for the systematic effects of artifacts and to obtain a realistic error bar, we performed the fitting both with and without applying the artifact mask. The resulting fits yield the H$\alpha$ line peak at \SI{22260(2)}{\AA} and \SI{22275(2)}{\AA}, corresponding to quasar redshifts of $2.3909 \pm 0.0003$ and $2.3931 \pm 0.0003$, respectively. For the final measurement, we adopted $z_{\mathrm{H}\alpha} = 2.3909$. The statistical uncertainty is $0.0003$, while the systematic uncertainty (defined as the difference 
between the two fits) is $0.0022$. Therefore, the total uncertainty is $\sigma_{\mathrm{total}} = \sqrt{\sigma_{\mathrm{stat}}^{2} + \sigma_{\mathrm{sys}}^{2}}=0.0023$ .

\citet{2020ApJ...893...14D} offer a method to correct the redshift measured from the blueshifted {\CIV} emission line to infer the systemic redshift, and the velocity offset are described by eq.~\eqref{eq:delta_v}:
\begin{equation}\label{eq:delta_v}
\begin{aligned}
\Delta v\left(\mathrm{~km} \mathrm{~s}^{-1}\right)= & \alpha \log _{10}\left(\mathrm{FWHM}_{{\CIV}}\right) \\
& +\beta \log _{10}\left(\mathrm{REW}_{\CIV}\right)+\gamma \log _{10}\left(L_{1350}\right),
\end{aligned}
\end{equation}
in which $\Delta v$ is the velocity shift between the measured redshift and the systemic redshift, $\mathrm{FWHM}_{{\CIV}}$ is the full width half maximum (FWHM) of {\CIV}, $\mathrm{REW}_{\CIV}$ is the rest-frame equivalent width (REW) of the {\CIV} line, and $L_{1350}$ is the rest-frame monochromatic luminosity at \SI{1350}{\AA}. For {\CIV}, the coefficients $\alpha = -3670\pm549$, $\beta = 1604\pm450$, and $\gamma = 217\pm48$, according to \citet{2020ApJ...893...14D}. Moreover, the systemic redshift can be estimated by eq.~\eqref{eq:systemic_z}:
\begin{equation}\label{eq:systemic_z}
z_{\text {sys}}=\frac{c \cdot z_{\text {meas}}-\Delta v}{\Delta v+c},
\end{equation}
where $z_{\text {sys}}$ and $z_{\text {meas}}$ represent the systemic redshift and measured redshift, respectively, $\Delta v$ is the velocity offset from eq.~\eqref{eq:delta_v}, and $c$ is the speed of light.

We fit a multi-Gaussian function to the {\CIV} emission line, from which we obtained a FWHM of \SI{4750(62)}{\km\per\s}, a REW of \SI{27(1)}{\AA}, and a measured redshift of $z_{\text {meas}} = 2.3841\pm0.0002$. In addition, $\log _{10}\left(L_{1350}\right)$ was measured to be \SI{45.24(0.02)} (with luminosity expressed in erg per second), calculated using the median flux in the rest-frame range 1330--\SI{1370}{\AA}. Based on eq.~\eqref{eq:delta_v} and \eqref{eq:systemic_z}, the systemic redshift was estimated to be $z_{\CIV} = 2.40\pm0.03$.

From the perspective of measurement errors, the redshift derived from H$\alpha$, $z_{\mathrm{H} \alpha} = 2.3908\pm0.0003$, is more accurate and was therefore adopted as our suggested best value for $z_{\mathrm{sys}}$. The systemic redshift derived from \ion{C}{IV} is consistent with this adopted value.

Following the method of \citet{2010MNRAS.408.2128F}, we performed spectral point spread function (SPSF) subtraction on the 2D spectra at all position angles to search for the H$\alpha$, {\OII}~$\lambda\lambda 3726, 3729$, and {\OIII}~$\lambda 5007$ emission lines from the host galaxy. Such lines were previously detected from DLA-galaxy counterparts \citep[e.g.,][]{2010MNRAS.408.2128F}. None of these emission lines were detected in any of the individual spectra nor in the combined spectrum. We also performed a 2-pixel binning of the 2D spectra to search for a possible Ly$\alpha$ emission line embedded within the Ly$\alpha$ absorption trough. No such emission feature was detected.

In Table \ref{tab:lines}, we list all absorption lines of the proximate DLA and measure their equivalent widths (EWs). Based on these absorption lines, we determined the precise redshift of the absorber to be $z = 2.4007 \pm 0.0003$. Compared to the redshift of the quasar, this is redshifted by 864 km s$^{-1}$. In our calculations, we did not use the {\CI} absorption lines, as they include fine-structure transitions. We also excluded absorption lines that are too close to each other and therefore blended.

In addition to the proximate DLA, we identified six other intervening systems at different redshifts. These systems, ordered by increasing redshift, are located at z = 1.7116,
1.7338, 1.7466, 1.7474, 1.8273, and 2.1859, with two, two, ten, three, five, and two identified absorption lines, respectively. The number of absorption systems in this sight line is significantly higher than what is typically observed toward quasars. A more extreme case was reported by \citet{2002ApJ...567L..13R}, where the quasar FIRST~0747$+$2739 was found to host 14 independent \ion{C}{iv} absorption systems. In \citet{2019MNRAS.488.5916S}, the large-sample analysis suggests that such an overdensity of narrow, intrinsic \ion{C}{iv} absorption lines is predominantly caused by outflows driven by accretion disk winds.

After using the VoigtFit package developed by \citet{2018arXiv180301187K}  to fit the Voigt profiles to the low-ionization metal lines, we noticed that the \ion{C}{II}~$\lambda1334$ line exhibits additional broadening. We interpret this as evidence for the presence of the fine-structure transitions \ion{C}{II}*~$\lambda1335.6$ and $\lambda1335.7$. According to \citet{2003ApJ...593..235W,2003ApJ...593..215W}, \ion{C}{II}*~$\lambda1335.7$ can be used to measure the star formation rate (SFR) in DLAs. However, due to the proximity of the quasar, the pumping of the fine structure lines could also be caused by emission from the quasar itself.  

\begin{table}[!t]
\centering
\begin{minipage}{0.5\textwidth}
\centering
\caption{Absorption lines of the $z = 2.4007$ DLA.}
\begin{tabular}{lll}
\noalign{\smallskip} \hline \hline \noalign{\smallskip}
Ion  & EW  & Redshift\\
     & (\AA) &    \\
\hline
\ion{N}{V}~$\lambda~1238$  &  $1.27 \pm 0.05$ & 2.4001 \\
\ion{N}{V}~$\lambda~1242$  &  $1.02 \pm 0.06$ &  2.3998\\
\ion{S}{II}~$\lambda~1250$ $^{(a)}$  &  $4.68 \pm 0.08$ & 2.4002 \\
\ion{S}{II}~$\lambda~1253$ $^{(a)}$  &  $4.42 \pm 0.07$ & 2.3986 \\
\ion{S}{II}~$\lambda~1259$  &  $0.99 \pm 0.06$ & 2.3992 \\
\ion{Si}{II}, \ion{Fe}{II}, \CI~$\lambda~1260$  &  $7.28 \pm 0.07$ & 2.3996\\
\CI~$\lambda~1277$  &  $1.03 \pm 0.02$ & 2.4000 \\
\CI~$\lambda~1280$  &  $0.56 \pm 0.06$ & 2.3998 \\
\ion{O}{I}~$\lambda~1302$  &  $6.41 \pm 0.07$ & 2.4007 \\ 
\ion{Si}{II}~$\lambda~1304$  &  $5.56 \pm 0.05$ & 2.4005 \\
\ion{Ni}{II}~$\lambda~1317$  &  $0.72 \pm 0.06$ & 2.3999 \\
\CI~$\lambda~1328$  &  $1.18 \pm 0.07$ & 2.3998 \\
\ion{C}{II}~$\lambda~1334$  &  $7.95 \pm 0.09$ & 2.4011\\ 
\ion{Cl}{I}~$\lambda~1347$  &  $3.26 \pm 0.07$ & 2.4062\\
\ion{Ni}{II}~$\lambda~1370$  &  $0.89 \pm 0.05$ & 2.4002\\
\ion{Si}{IV}~$\lambda~1393$  &  $4.61 \pm 0.05$ & 2.4003\\
\ion{Si}{IV}~$\lambda~1402$  &  $3.88 \pm  0.06$ & 2.4001\\
\ion{Ni}{II}~$\lambda~1454$  &  $0.23 \pm 0.04$ & 2.3960\\
\ion{Ni}{II}~$\lambda~1467$  &  $0.28 \pm 0.04$ & 2.4004\\
\ion{Si}{II}~$\lambda~1526$  &  $6.72 \pm 0.04$ & 2.4006\\
\ion{C}{IV}~$\lambda~1548$  &  $5.31 \pm 0.03$ & 2.4002\\
\ion{C}{IV}~$\lambda~1550$  &  $4.53 \pm 0.03$ & 2.4000\\
\CI~$\lambda~1560$  &  $1.22 \pm 0.04$ & 2.4004 \\
\ion{Fe}{II}~$\lambda~1608$  &  $1.10 \pm 0.04$ & 2.3990 \\
\ion{Fe}{II}~$\lambda~1611$  &  $3.62 \pm 0.04$ & 2.3952\\
\CI~$\lambda~1656$  &  $3.62 \pm 0.04$ & 2.4002 \\
\ion{Al}{II}~$\lambda~1670$  &  $7.04 \pm 0.11 $ & 2.4007 \\
\ion{Ni}{II}~$\lambda~1741$  &  $0.98 \pm 0.10$ & 2.4006 \\
\ion{Ni}{II}~$\lambda~1751$  &  $0.79 \pm 0.15$ & 2.4008 \\
\ion{Si}{II}~$\lambda~1808$  &  $2.36 \pm 0.09$ & 2.4003 \\
\ion{Al}{III}~$\lambda~1854$  &  $4.79 \pm 0.09$ & 2.4005 \\
\ion{Al}{III}~$\lambda~1862$  &  $3.51 \pm  0.08$ & 2.4003 \\
\ion{Cr}{II}, \ion{Zn}{II}~$\lambda~2026$  &  $2.98 \pm 0.07$ & 2.4002 \\
\ion{Cr}{II}~$\lambda~2056$  &  $0.79 \pm 0.07$ & 2.4004 \\
\ion{Cr}{II}, \ion{Zn}{II}~$\lambda~2062$  &  $1.37 \pm 0.07$ & 2.4003 \\
\ion{Cr}{II}~$\lambda~2066$  &  $0.15 \pm 0.04$ & 2.3992 \\
\ion{C}{II}~$\lambda~2325$  &  $0.69 \pm 0.04$ & 2.3996 \\
\ion{Fe}{II}~$\lambda~2344$  &  $9.13 \pm 0.07$ & 2.4006 \\
\ion{Fe}{II}~$\lambda~2374$  &  $6.85 \pm 0.05$ & 2.4006 \\
\ion{Fe}{II}~$\lambda~2382$  &  $10.91 \pm 0.06$ & 2.4007 \\
\ion{Mn}{II}~$\lambda~2576$  &  $1.03 \pm 0.10$ & 2.4005 \\
\ion{Fe}{II}~$\lambda~2586$  &  $9.65 \pm  0.07$ & 2.4006 \\
\ion{Mn}{II}~$\lambda~2594$  &  $0.65 \pm 0.03$ & 2.4006 \\
\ion{Fe}{II}~$\lambda~2600$  &  $11.95 \pm 0.12$ & 2.4007 \\
\ion{Mn}{II}~$\lambda~2606$  &  $0.56 \pm 0.04$ & 2.4006 \\
\ion{Mg}{II}~$\lambda~2796$  &  $14.22 \pm 0.08$ & 2.4017 \\
\ion{Mg}{II}~$\lambda~2796$  &  $13.51 \pm 0.05$ & 2.4017 \\
\ion{Mg}{I}~$\lambda~2796$  &  $6.82\pm 0.12$ & 2.4012 \\
\hline
\noalign{\smallskip} \hline \noalign{\smallskip}
\end{tabular}
\vspace{2mm}
\parbox{\linewidth}{\footnotesize \textbf{Note.} $^{(a)}$\ion{S}{II}~$\lambda~1250$ and \ion{S}{II}~$\lambda~1253$ are blended with \ion{C}{IV}~$\lambda~1548$ and \ion{C}{IV}~$\lambda~1550$ at z=1.747, respectively.}
\label{tab:lines}
\end{minipage}
\end{table}

\subsection{Extinction}

In the extinction curve, we found that our DLA does not exhibit a prominent 2175 \AA\ bump. According to \citet{2003ApJ...594..279G}, the Small Magellanic Cloud (SMC) extinction curve similarly shows little to no evidence of the bump, in contrast to the prominent feature observed in the Milky Way (MW) and Large Magellanic Cloud (LMC). Therefore, we adopted the Fitzpatric \& Massa (FM) parameters, i.e., the extinction parameters from \citet{2007ApJ...663..320F}, corresponding to the SMC-type extinction curve. Then we performed a Markov chain Monte Carlo (MCMC) analysis to determine the optimal values of $E(B-V)$ and $R_V$ that minimize the $\chi^2$ between the reddened quasar template and the observed spectrum. The best-fit parameters are $E(B-V) = 0.3995 \pm 0.0006$ (this error bar is the formal error resulting from the fit, but we consider it unrealistically small) and $R_V = 2.8 \pm 0.5$. Based on these parameters, the reddened quasar template is plotted in Fig. \ref{Spectrum}. Using the dust maps \citep{1998ApJ...500..525S,2011ApJ...737..103S}, we obtained a Milky Way reddening contribution of $E(B-V)=0.0131 \pm 0.0006$. This value is much lower than the extinction inferred for the absorber and can therefore be neglected.

During the MCMC fitting, we also included the Lyman-alpha absorption component in the reddened model to determine the neutral hydrogen column density of the DLA, resulting in $\log N(\ion{H}{I}) = 21.214 \pm 0.003$ (this error bar is the formal error resulting from the fit, but we consider it unrealistically small). The Lyman-$\alpha$ absorption line was approximated using the Voigt-Hjerting function following the approximation of \citet{Tepper_Garc_a_2006}. 

\subsection{Metallicity}
We performed the Voigt profile fitting for all singly ionized metal lines. In the fitting process, we applied the resolution of the three arms, determined by measuring the Gaussian FWHM along the spatial axis of the 2D spectrum and the linear relation given by \citet{2019A&A...623A..92S}. Based on the column densities of the various metals obtained from the Voigt profile fitting and the photospheric solar abundances provided by \citet{2021A&A...653A.141A}, we calculated the metallicity for each individual element: [Zn/H] $= -0.22 \pm 0.04$, [Fe/H] $> -1.16 \pm 0.08$, [Cr/H] $= -1.08 \pm 0.08$, [Si/H] $> -0.517 \pm 0.05$, [Mn/H] $= -1.22 \pm 0.06$, [Ni/H] $= -1.04 \pm 0.06$, [S/H]$= 0.01 \pm 0.54$, and [Al/H]$> -0.87 \pm 0.33$. Hence, the system can be classified as a metal-rich DLA.

\begin{figure}[th]
\centering
\includegraphics[scale=0.5]{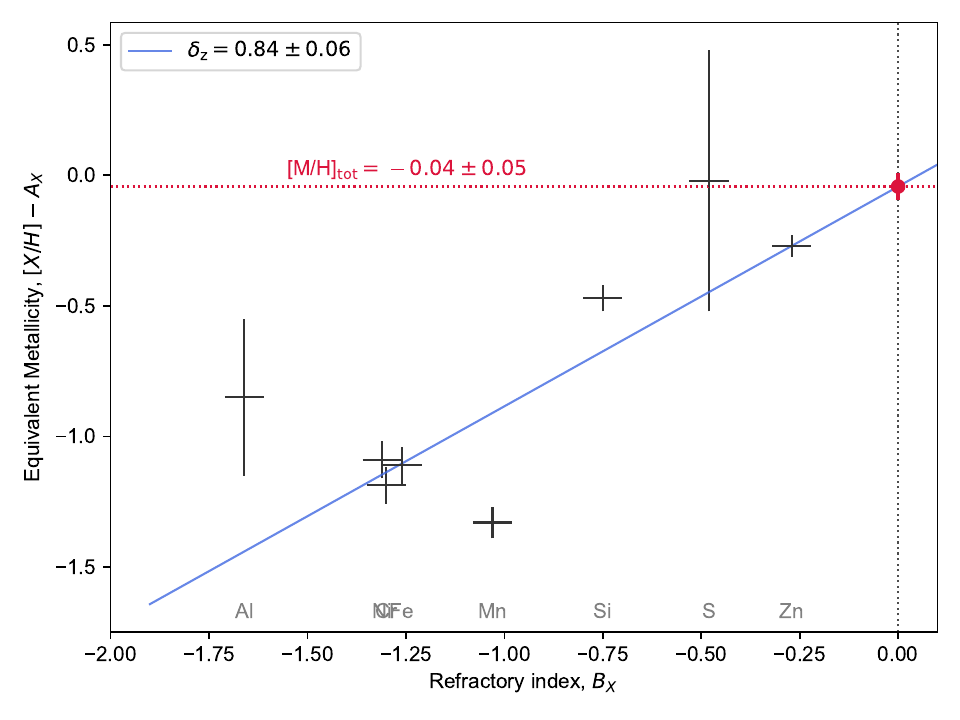}
\caption{Metallicities of the system corrected for dust using the refractory index $B_X$, with $B_X$ values for all elements taken from Table~3 of \citet{2016A&A...596A..97D} and Table~1 of \citet{2024A&A...691A.129K}. The blue curve represents a linear fit to Fe and Zn, with the slope $\delta_Z$ indicating dust depletion, and the intercept $M_{\rm tot}$ representing the corrected total metallicity of the system. Notably, Al and S lie above the curve, which can be attributed to $\alpha$-enrichment.}
\label{fig:correction}
\end{figure}

Most elements deplete onto dust grains at some level. Among all the measured metallicities, the metallicity of Zn is the closest to the metallicity of the system as Zn exists predominantly in the gas phase \citep{1990ApJ...363...57M,1990AuJPh..43..227P,1994ApJ...426...79P}. To obtain a more accurate estimate of the metallicity, we applied the dust depletion correction methods of \citet{2016A&A...596A..97D}. In Figure~\ref{fig:correction}, we present the metallicity of all elements versus the refractory index. A linear fit using Fe and Zn yields a slope of $\delta_Z = 0.84$, which represents a moderate level of dust depletion. The intercept of this curve at the origin of the $B_X$ axis corresponds to the corrected total abundance, $M_{\rm tot}$. The intercept of this curve at $B_X=0$ corresponds to the corrected total abundance, $M_{\rm tot} = -0.04 \pm 0.05$, indicating that the total metallicity of the system is consistent with the solar value. Meanwhile, due to the contribution from nucleosynthesis, some elements exhibit over- or underabundances beyond dust depletion. In Figure~\ref{fig:correction}, we note an $\alpha$-enrichment of $\sim 0.2$~dex in the metallicities of S and Si, indicating recent star formation activity \citep{2011MNRAS.418L..74D}.

Using the metallicity, we also derived the metals-to-dust ratio of the system to be \(20.95 \pm 0.09\) by $\left(\log \frac{N_{\ion{H}{I}}}{\rm cm^{-2}} + [{\rm M/H}]\right) - \log \frac{A_V}{\rm mag}$. According to the study of \citet{2013A&A...560A..26Z} of a sample that spans redshifts \(z = 0.1\)--\(6.3\), the metals-to-dust ratio in quasar absorbers is approximately constant at a value of 21.2 within this redshift range. A more recent study by \citet{2023A&A...679A..91H} that focused on \(z = 0.1\)--\(6.3\) found a similar value of 21.4. Our result is in good agreement with these values.

\begin{figure}[th]
\centering
\includegraphics[scale=0.27]{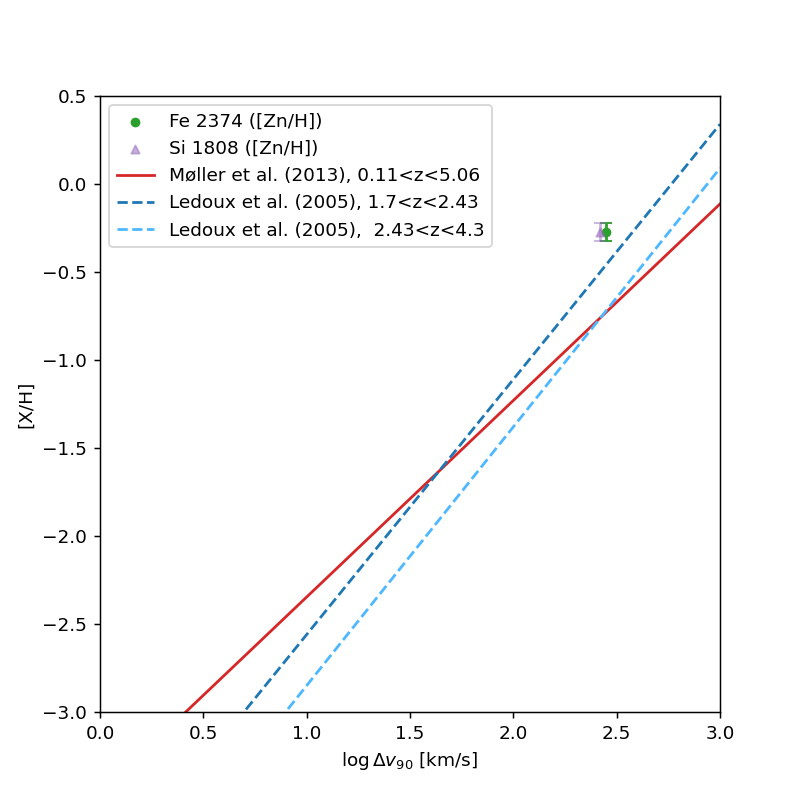}
\caption{Mass–metallicity relation of Q\,2310$-$3358 derived through the correlation between logarithmic velocity width and metallicity, measured by two unsaturated metal lines, \ion{Fe}{II}~$\lambda2374$ and \ion{Si}{II}~$\lambda1808$. The two dashed blue lines show the linear relation fit by \citet{2006A&A...457...71L} based on 70 DLAs in the redshift range \(1.7 < z < 4.3\), while the solid red line illustrates the linear relation from \citet{2013MNRAS.430.2680M}, based on 110 DLAs spanning \(0.11 < z < 5.06\).}
\label{mass_metallicity}
\end{figure}

Taking into account the dynamical properties of the system, we selected low-ionization metal lines that are neither saturated nor too weak, to analyze the velocity width of the system.  
\ion{Si}{II}~$\lambda~1808$ and \ion{Fe}{II}~$\lambda~2374$ proved to be excellent candidates, with measured velocity widths of $264\,\mathrm{km\,s^{-1}}$ and $282\,\mathrm{km\,s^{-1}}$, respectively. 
We adopted the metallicity of the system traced by zinc, $\mathrm{[Zn/H]}=-0.22\pm0.04$, and placed the two measured velocity widths in the $\Delta v_{90}$ versus metallicity diagram, as shown in Fig.~\ref{mass_metallicity}. The mass-metallicity relations derived from the large sample studies by \citet{2013MNRAS.430.2680M} and \citet{2006A&A...457...71L} are also plotted for comparison. We observe that Q\,2310$-$3358 exhibits a slightly elevated metallicity relative to the mass--metallicity relation while still following the overall trend.

\subsection{Neutral carbon and molecules}
Neutral carbon has a low ionization potential of 11.26\,eV, and it is therefore not shielded by neutral hydrogen in the ISM. According to \citet{2018A&A...612A..58N}, diffuse molecular gas such as H$_2$ and CO is often observed with {\CI}. In these regions, HI and dust together create a self-shielding regime: HI scatters the UV radiation, while dust attenuates the ionizing photons, thereby allowing the survival of these atoms and molecules.

\begin{figure*}[th]
\centering
\includegraphics[scale=0.84]{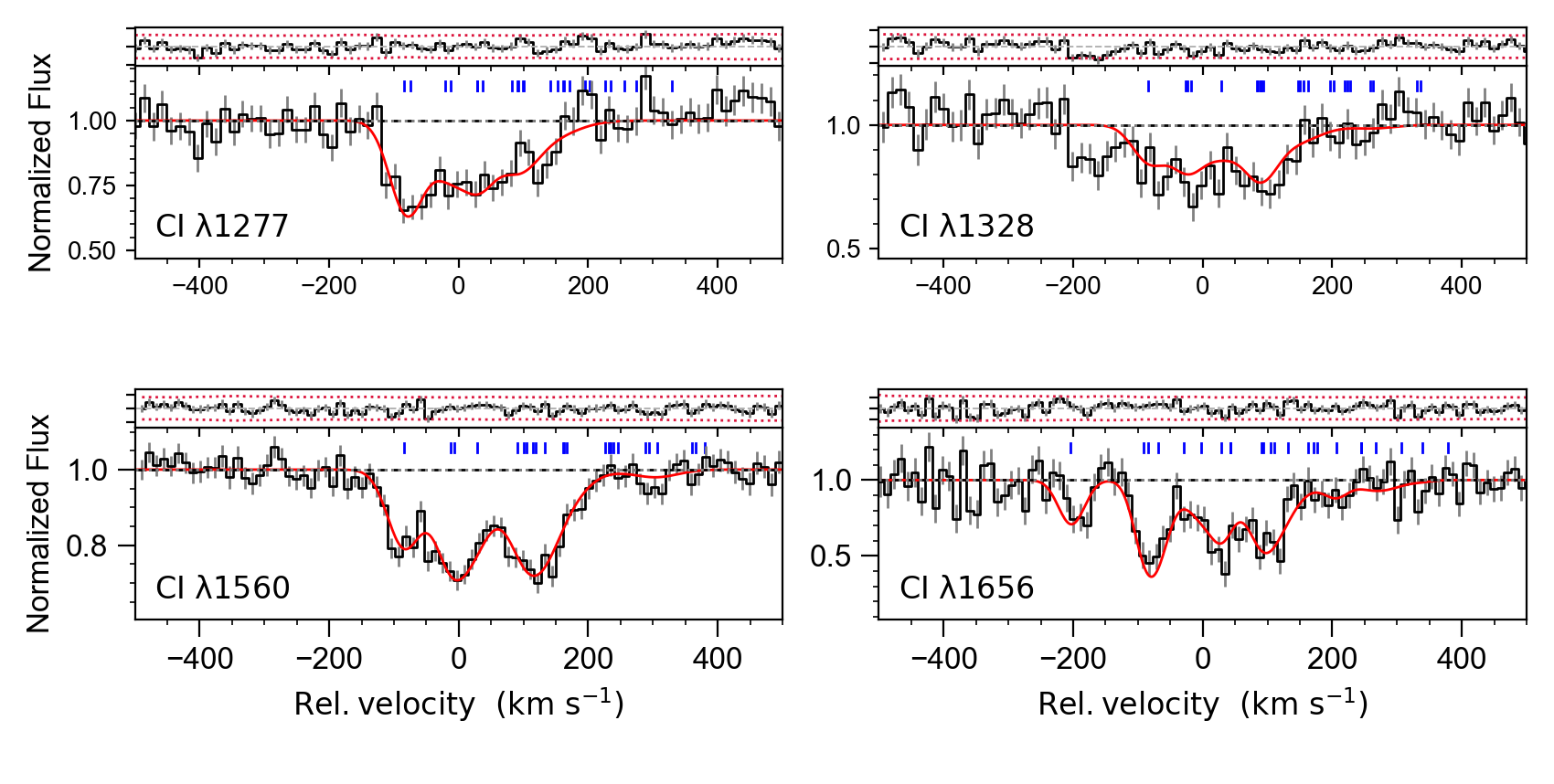}
\caption{Voigt profile fitting of four multiplets of {\CI}. The spectrum is shown as the black line with the best-fitting model overlaid as the solid red curve. Each of the blue tick marks indicates the position of a velocity component of either the ground level ($J=0$) or the two excited fine-structure levels ($J=1$ and $J=2$). The small panel on top of each sub-figure shows the residuals of the fit in the given region.}
\label{fig:CI}
\end{figure*}

We performed Voigt profile fitting for {\CI}~$\lambda~1277$, {\CI}~$\lambda~1328$, {\CI~}$\lambda~1560$, {\CI}~$\lambda~1656$, and their fine-structure levels with $J=0,1,2$. The results, shown in Figure~\ref{fig:CI}, clearly indicate the presence of {\CI}, {\CI}*, and {\CI}**. The column density of {\CI} is $\log N(\ion{C}{I}) = 14.08 \pm 0.09$, while the column densities of the fine-structure levels $J=1,2$, {\CI}* and {\CI}**, are $14.050 \pm 0.078$ and $13.623 \pm 0.103$, respectively. Such a strong DLA exhibiting {\CI} absorption is very rare among quasar absorbers, as the main population of {\CI} absorbers has been found in sub-DLAs with $N(\ion{H}{I})=2\times10^{20}$ \citep{2015A&A...580A...8L}. In addition, {\CI} absorption is more frequently detected in gamma-ray burst (GRB) host galaxies than in quasar absorption systems, as it also tends to arise in environments with relatively high metallicity and dust content \citep{2019A&A...621A..20H,2019A&A...629A.131H,2020ApJ...889L...7H}.

\begin{figure*}[th]
    \centering
    \includegraphics[scale=0.72]{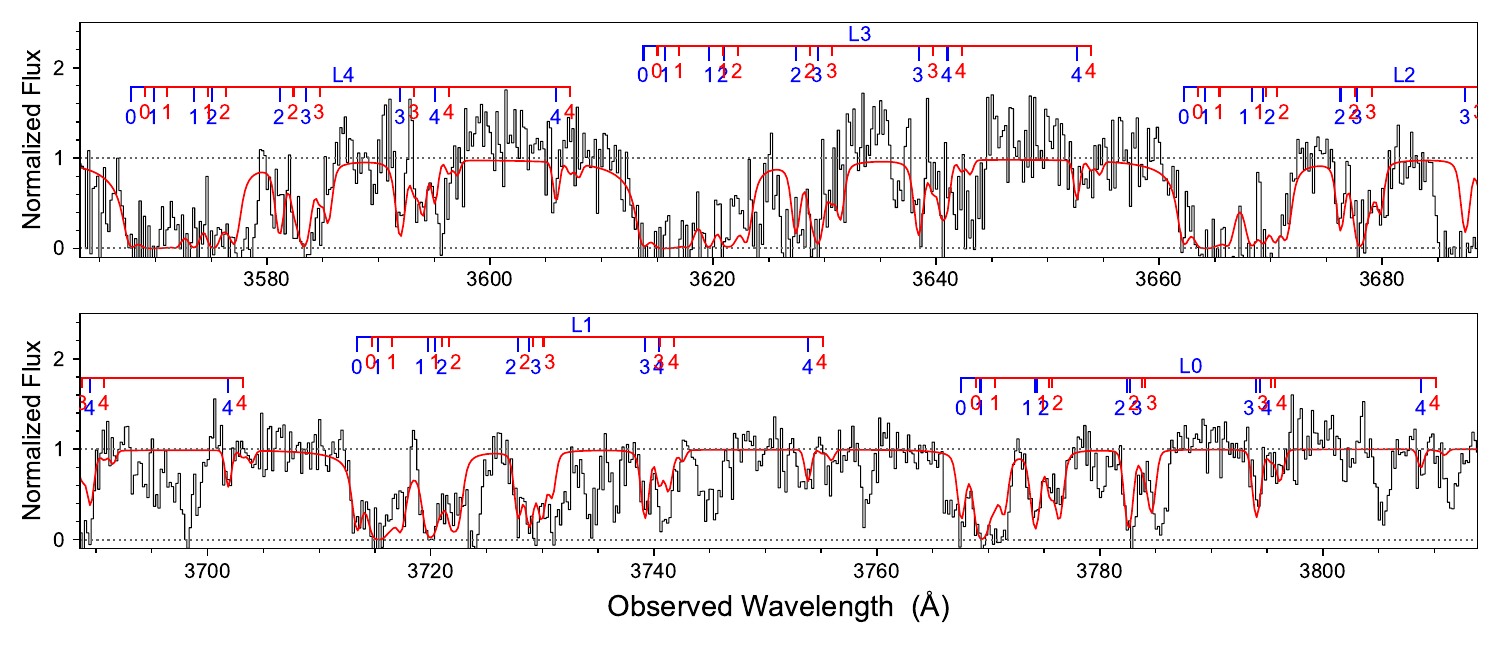}
    \caption{Voigt profile fitting of H$_2$ lines from rotational and vibrational levels. The red line shows the best-fit model using VoigtFit, and the blue tick marks show the location of the various rotational transitions (0--4) of each of the vibrational bands ($\nu$) to the ground state denoted as $BX(\nu-0)$. The model assumes two velocity components.}
    \label{fig:H2fit}
\end{figure*}

Due to the self-shielding regime, the successful detection of strong {\CI} absorption often implies the coexistence of H$_2$ and CO molecules in the DLA \citep{1981ApJ...246L.147L, 2005MNRAS.362..549S,2018A&A...612A..58N}. To characterize the molecular hydrogen absorption, we fit the H$_2$ lines from different vibrational and rotational levels using Voigt profile fitting, under the assumption of two velocity components, as shown in Fig.~\ref{fig:H2fit}. In this framework, the H$_2$ transitions are denoted as $BX(\nu-0)$ for the Lyman band, which corresponds to those from the vibrational level $\nu$ of the excited $B$ state to the vibrational ground level of the $X$ state. 

The population distribution of the rotational excitation of molecular hydrogen can be described using the Boltzmann distribution by eq.~\eqref{molecularH}:
\begin{equation}\label{molecularH}
\frac{N(H_2,J')}{g(H_2,J')}=\frac{N(H_2,J)}{g(H_2,J)}e^{-E_{J'J}/kT_{J'J}},
\end{equation}
where $g(\rm H_2,J)$ and $g(\rm H_2,J')$ are the degeneracies of the $J$ and $J'$ levels, respectively, while $N(\rm H_2,J)$ and $N(\rm H_2,J')$ represent the corresponding column densities.

\begin{figure}[th]
\centering
\includegraphics[scale=0.45]{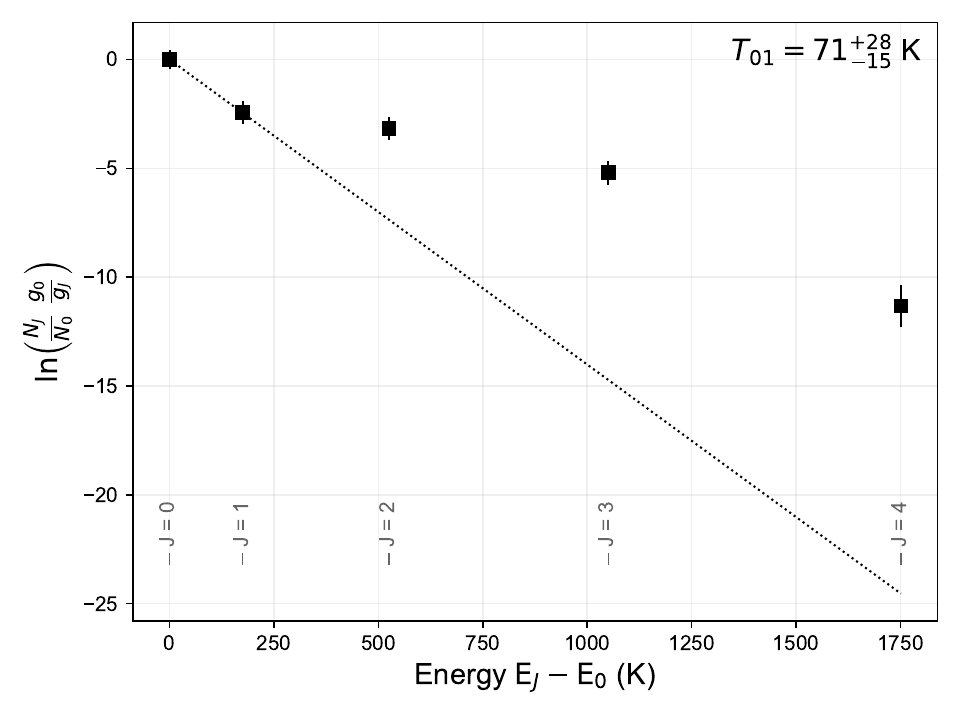}
\caption{Excitation diagram of H$_2$ showing the population of rotational levels relative to the ground state. The straight dotted line shows the best-fit model of the excitation temperature derived from the first two $J$-levels assuming a Boltzmann distribution. Higher order levels indicate higher excitation, which is commonly observed \citep{Noterdaeme2017}.}
\label{fig:excitation}
\end{figure}

In logarithmic form and for $J=0$, Eq.~\eqref{molecularH} can be written as a linear relation of $\ln(N_J/N_0 \, g_0/g_J)$ versus $E_J - E_0$, with a slope of $-1/kT$. At high column densities, $T_{01}$ is often used to represent the kinetic temperature of the H$_2$ gas \citep{2006MNRAS.365L...1R}. Therefore, we performed a linear fit for the $J=0$ and $J=1$ levels (see Fig. \ref{fig:excitation}), obtaining an excitation temperature of $T_{01} = 71^{+28}_{-15}~\mathrm{K}$. This is a relatively low excitation temperature, significantly lower than the typical values of $T_{01} \sim 130$--$160$~K in low-redshift and high-redshift DLA samples \citep{2005MNRAS.362..549S,2015MNRAS.448.2840M}, but very close to 77~K in the Galactic disk and 71~K in the Magellanic Clouds (MCs) \citep{1977ApJ...216..291S,2002AAS...201.7702R,2012ApJ...745..173W}.

In addition, we find excess excitation for the $J\geq2$ rotational levels. A stronger excitation of the higher $J$-levels is commonly observed as previously reported \citep{Noterdaeme2017}. However, in our case, the excess is particularly pronounced even for $J=2$, which could indicate photon pumping by a strong UV radiation field.

\begin{figure}[th]
\centering
\includegraphics[scale=0.5]{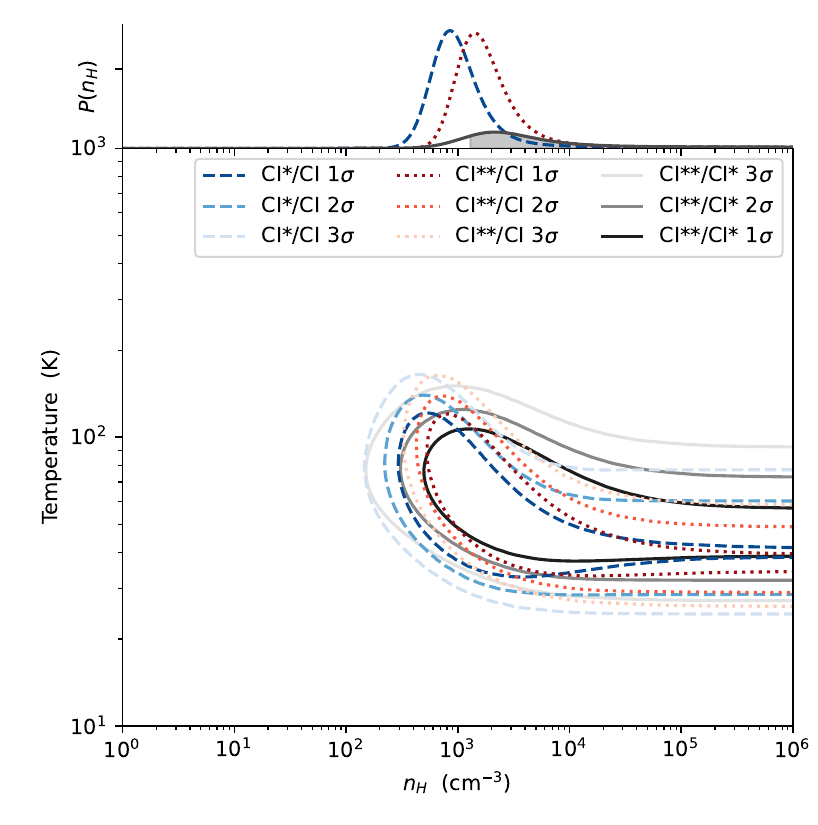}
\caption{Constraints on temperature vs. hydrogen density from the relative population of {\CI} fine-structure levels. The contours show 1 to 3 sigma confidence regions (dark to light shading) from three different line ratios (J=1 to J=0, dashed blue; J=2 to J=0, dotted red; and J=2 to J=1, solid black). All constraints assume an excitation temperature of 71 K derived from H$_2$ rotational levels.}
\label{fig:CI_density}
\end{figure}

Figure~\ref{fig:CI_density} shows the constraints on the physical conditions of the absorbing gas obtained from the relative populations of the \ion{C}{I} fine-structure levels. The confidence regions (1--3$\sigma$) are derived independently from three line ratios of $J=0,1,2$. Each ratio provides a distinct locus in the temperature--density plane, reflecting the different sensitivities of the fine-structure transitions to collisional excitation \citep{2010ApJ...722..460J}. The overlapping region of the three constraints defines the most probable range of hydrogen density and kinetic temperature. In the present analysis, we assume a strong UV radiation field and excitation temperature of $T_{\rm ex}=71$\,K, as inferred from the H$_2$ rotational levels. The resulting joint constraints favor a cold and fairly high density phase, with an $n_{H}$ of $\sim1000$\,cm$^{-3}$. If the ambient UV radiation field is weaker or the kinetic temperature is higher, the required density could be slightly lower, down to around $600\,\mathrm{cm}^{-3}$. Moreover, We derived an upper limit on the CO column density, finding $\log N(\mathrm{CO}) < 13.9$, which is consistent with the measured {\CI} column density of $\sim14.3$, following the relation reported by \citet{2018A&A...612A..58N} for intervening absorbers.

\section{Discussion and conclusions} 
\label{sec:discussion and conclusions}

Q\,2310$-$3358 is a dust-reddened quasar with a strong, metal-rich proximate DLA. It was selected using a specific Gaia-based astrometric method combined with a photometric selection rule, which has been shown to efficiently select hitherto overlooked reddened quasars -- in this case a quasar reddened by a proximate dusty DLA. By comparing with Figure~9 of \citet{2023A&A...673A..89N}, we find that Q\,2310--3358, similar to other known proximate molecular absorbers, lies at the upper edge of the expected dust-bias trend \citep{1998A&A...333..841B}. Since the sample of \citet{2023A&A...673A..89N} was optically selected from SDSS, the presence of dust bias may have caused many such absorbers to be overlooked.

In several aspects, Q\,2310$-$3358 follows the correlations observed for intervening DLAs. It is consistent with the approximately constant metal-to-dust ratio found in large-sample studies \citep{2013A&A...560A..26Z,2023A&A...679A..91H} and also aligns with previous results on the mass--metallicity relation \citep{2006A&A...457...71L,2013MNRAS.430.2680M}.

Using the H$\alpha$ and the corrected \ion{C}{IV} emission lines, we obtained quasar redshifts of $z_{em} = 2.3909 \pm 0.0022$ and $z_{em} = 2.40 \pm 0.03$, respectively. The absorber has a redshift of $z_{abs} = 2.4007 \pm 0.0003$, i.e., redshifted by 864 km s$^{-1}$ relative to the quasar. Such large relative velocities are common in proximate DLAs \citep{Ellison2010}. In \cite{2023A&A...673A..89N}, Fig.~11 presents 13 proximate molecular quasar absorbers with velocities relative to their quasars, showing an average redshift of $250~\mathrm{km~s}^{-1}$, which is significantly lower than the value we measured. Only a very small number of systems reach relative velocities comparable to ours.

If we adopt the redshift of the quasar measured from the H$\alpha$ emission line and consider the precisely determined absorber redshift of $z_{\mathrm{abs}} = 2.4007$, this quasar absorption-line system represents a peculiar case where $z_{\mathrm{em}} < z_{\mathrm{abs}}$. How such a scenario can occur was already discussed by \citet{1978PhyS...17..217W}. The relative redshifts imply that the absorption system and the quasar are moving toward one another. Given its very high relative velocity (875~km~s$^{-1}$), this suggests that the absorber is either a foreground galaxy with a large peculiar velocity within the quasar group environment, or gas in-falling toward the central engine at high velocity, as typical galactic motions are on the order of a few hundred kilometers per second. Considering the other indications of very high metallicity, strong excitation of H$_2$, and high density inferred from \ion{C}{I}, we regard the latter scenario as more plausible. Altogether, these properties could arise in a highly disturbed host galaxy following a merger event, which would both trigger active galactic nucleus (AGN) activity and induce a strong starburst, naturally explaining the observed $\alpha$-enhancement.
 
Considering that the excess excitation at the $J=2$ rotational level implies the influence of photon pumping by a strong UV background, we infer that the quasar and the DLA are likely to be in very close proximity. Also from this perspective, the second explanation appears more plausible.

The detection of {\CI} and its fine-structure levels is important for understanding the physical conditions within molecular clouds. By analyzing the three {\CI} levels, we constrained the excitation temperature to be $T_{\mathrm{01}} \sim 71\,\mathrm{K}$ and the gas density of the system to be $n_{\mathrm{H}}\sim 1000\,\mathrm{cm}^{-3}$. These conditions are essential for the survival of H$_2$ and CO. An upper limit was derived for CO ($\log N(\mathrm{CO}) < 13.9$). The coexistence of CO with H$_2$ and {\CI} supports the presence of a self-shielded regime \citep{2018A&A...612A..58N}.

Clear detections of {\CI} and H$_2$ are very rare \citep{2015A&A...580A...8L}. The  system exhibits significant metal enrichment ([M/H]$_{tot}=-0.04$) and strong dust reddening. Such DLAs are rarely observed, and this work provides a new valuable example of this rare type of absorption systems. Similarly, previous studies of another absorber with $z_{\mathrm{em}} \simeq z_{\mathrm{abs}}$, Q0528$-$250, have also revealed the presence of {\CI} and H$_2$, which allowed constraints to be placed on the physical conditions. \citet{1998A&A...335...33S} reported the first detection of H$_2$ in Q0528$-$250 and demonstrated that the presence of dust plays a crucial role in facilitating H$_2$ formation. \citet{2020MNRAS.497.1946B} further investigated this DLA and estimated the distance between the quasar and the absorbing cloud, which constrained the physical conditions of the gas in Q0528$-$250 to an excitation temperature of $\sim150$~K and a hydrogen number density of $n_{\rm H} \sim 200~\mathrm{cm}^{-3}$. In contrast, our DLA exhibits a more peculiar environment with a lower temperature ($71$~K) and a higher density ($\sim 10^{3}~\mathrm{cm}^{-3}$), which may be related to the high metallicity of Q\,2310$-$3358. Another comparable case is the study by \citet{2025Natur.641.1137B} on J012555.11$-$012925.00, a merging galaxy system at $z = 2.7$. The companion galaxy exhibits DLA characteristics and is also classified as a proximate DLA. Compared to this system, Q\,2310$-$3358 shows a much higher metallicity, implying a more evolved and actively star-forming environment. J012555.11$-$012925.00 also exhibits a lower velocity of about 550$\mathrm{km~s}^{-1}$, further highlighting that our DLA has an unusually high redshift relative to the quasar.

\begin{acknowledgements}
The Cosmic Dawn Center (DAWN) is funded by the Danish National
Research Foundation under grant No. 140.  JPUF is supported by the
Independent Research Fund Denmark (DFF--4090-00079).  LC is supported by the Independent
Research Fund Denmark (DFF-2032-00071).
J.K.K. is supported by the French Agence Nationale de la Recherche (ANR) under grant number ANR-24-CE31-7454 (CI-CNM). This research has made use of the VizieR catalog access tool, CDS, Strasbourg Astronomical Observatory, France (DOI : 10.26093/cds/vizier).
\end{acknowledgements}

\bibliographystyle{aa}
\bibliography{ref}

@ARTICLE{1995Natur.375..469W,
       author = {{Webster}, Rachel L. and {Francis}, Paul J. and {Petersont}, Bruce A. and {Drinkwater}, Michael J. and {Masci}, Frank J.},
        title = "{Evidence for a large undetected population of dust-reddened quasars}",
      journal = {\nat},
         year = 1995,
        month = jun,
       volume = {375},
       number = {6531},
        pages = {469-471},
          doi = {10.1038/375469a0},
       adsurl = {https://ui.adsabs.harvard.edu/abs/1995Natur.375..469W}
}

@ARTICLE{Glikman2022,
       author = {{Glikman}, E. and {Lacy}, M. and {LaMassa}, S. and {Bradley}, C. and {Djorgovski}, S.~G. and {Urrutia}, T. and {Gates}, E.~L. and {Graham}, M.~J. and {Urry}, M. and {Yoon}, I.},
        title = "{The WISE-2MASS Survey: Red Quasars Into the Radio Quiet Regime}",
      journal = {\apj},
         year = 2022,
        month = aug,
       volume = {934},
       number = {2},
          eid = {119},
        pages = {119},
          doi = {10.3847/1538-4357/ac6bee},
archivePrefix = {arXiv},
       eprint = {2204.13745},
 primaryClass = {astro-ph.GA},
       adsurl = {https://ui.adsabs.harvard.edu/abs/2022ApJ...934..119G}
}

@ARTICLE{Glikman2013,
       author = {{Glikman}, Eilat and {Urrutia}, Tanya and {Lacy}, Mark and {Djorgovski}, S.~G. and {Urry}, Meg and {Croom}, Scott and {Schneider}, Donald P. and {Mahabal}, Ashish and {Graham}, Matthew and {Ge}, Jian},
        title = "{Dust Reddened Quasars in FIRST and UKIDSS: Beyond the Tip of the Iceberg}",
      journal = {\apj},
         year = 2013,
        month = dec,
       volume = {778},
       number = {2},
          eid = {127},
        pages = {127},
          doi = {10.1088/0004-637X/778/2/127},
archivePrefix = {arXiv},
       eprint = {1309.6626},
 primaryClass = {astro-ph.CO},
       adsurl = {https://ui.adsabs.harvard.edu/abs/2013ApJ...778..127G}
}

@ARTICLE{Glikman2018,
       author = {{Glikman}, E. and {Lacy}, M. and {LaMassa}, S. and {Stern}, D. and {Djorgovski}, S.~G. and {Graham}, M.~J. and {Urrutia}, T. and {Lovdal}, Larson and {Crnogorcevic}, M. and {Daniels-Koch}, H. and {Hundal}, Carol B. and {Urry}, M. and {Gates}, E.~L. and {Murray}, S.},
        title = "{Luminous WISE-selected Obscured, Unobscured, and Red Quasars in Stripe 82}",
      journal = {\apj},
         year = 2018,
        month = jul,
       volume = {861},
       number = {1},
          eid = {37},
        pages = {37},
          doi = {10.3847/1538-4357/aac5d8},
archivePrefix = {arXiv},
       eprint = {1805.06961},
 primaryClass = {astro-ph.GA},
       adsurl = {https://ui.adsabs.harvard.edu/abs/2018ApJ...861...37G}
}

@ARTICLE{Weymann1977,
       author = {{Weymann}, R.~J. and {Williams}, R.~E. and {Beaver}, E.~A. and {Miller}, J.~S.},
        title = "{A spectroscopic study of selected quasars with z$_{abs}$ > z$_{em}$.}",
      journal = {\apj},
         year = 1977,
        month = may,
       volume = {213},
        pages = {619-631},
          doi = {10.1086/155193},
       adsurl = {https://ui.adsabs.harvard.edu/abs/1977ApJ...213..619W}
}

@ARTICLE{Ellison2010,
       author = {{Ellison}, Sara L. and {Prochaska}, J. Xavier and {Hennawi}, Joseph and {Lopez}, Sebastian and {Usher}, Christopher and {Wolfe}, Arthur M. and {Russell}, David M. and {Benn}, Chris R.},
        title = "{The nature of proximate damped Lyman {\ensuremath{\alpha}} systems}",
      journal = {\mnras},
         year = 2010,
        month = aug,
       volume = {406},
       number = {3},
        pages = {1435-1459},
          doi = {10.1111/j.1365-2966.2010.16780.x},
archivePrefix = {arXiv},
       eprint = {1004.2715},
 primaryClass = {astro-ph.CO},
       adsurl = {https://ui.adsabs.harvard.edu/abs/2010MNRAS.406.1435E}
}

@ARTICLE{Noterdaeme2019,
       author = {{Noterdaeme}, P. and {Balashev}, S. and {Krogager}, J. -K. and {Srianand}, R. and {Fathivavsari}, H. and {Petitjean}, P. and {Ledoux}, C.},
        title = "{Proximate molecular quasar absorbers. Excess of damped H$_{2}$ systems at z$_{abs}$ {\ensuremath{\approx}} z$_{QSO}$ in SDSS DR14}",
      journal = {\aap},
         year = 2019,
        month = jul,
       volume = {627},
          eid = {A32},
        pages = {A32},
          doi = {10.1051/0004-6361/201935371},
archivePrefix = {arXiv},
       eprint = {1905.02040},
 primaryClass = {astro-ph.GA},
       adsurl = {https://ui.adsabs.harvard.edu/abs/2019A&A...627A..32N}
}

@ARTICLE{Moller1998,
       author = {{M{\o}ller}, P. and {Warren}, S.~J. and {Fynbo}, J.~U.},
        title = "{On the nature of z\_abs \raisebox{-0.5ex}\textasciitilde z\_em damped absorbers in quasar spectra}",
      journal = {\aap},
         year = 1998,
        month = feb,
       volume = {330},
        pages = {19-24},
          doi = {10.48550/arXiv.astro-ph/9709160},
archivePrefix = {arXiv},
       eprint = {astro-ph/9709160},
 primaryClass = {astro-ph},
       adsurl = {https://ui.adsabs.harvard.edu/abs/1998A&A...330...19M}
}

@ARTICLE{1998A&A...335...33S,
       author = {{Srianand}, R. and {Petitjean}, Patrick},
        title = "{Molecules in the z\_abs = 2.8112 damped system toward PKS 0528-250}",
      journal = {\aap},
         year = 1998,
        month = jul,
       volume = {335},
        pages = {33-40},
          doi = {10.48550/arXiv.astro-ph/9804036},
archivePrefix = {arXiv},
       eprint = {astro-ph/9804036},
 primaryClass = {astro-ph},
       adsurl = {https://ui.adsabs.harvard.edu/abs/1998A&A...335...33S}
}

@ARTICLE{2020A&A...644A..17H,
       author = {{Heintz}, K.~E. and {Fynbo}, J.~P.~U. and {Geier}, S.~J. and {M{\o}ller}, P. and {Krogager}, J. -K. and {Konstantopoulou}, C. and {de Burgos}, A. and {Christensen}, L. and {Steinhardt}, C.~L. and {Milvang-Jensen}, B. and {Jakobsson}, P. and {H{\o}g}, E. and {Arvedlund}, B.~E.~H.~K. and {Christiansen}, C.~R. and {Hansen}, T.~B. and {Henriksen}, P.~D. and {Kuszon}, K.~B. and {McKenzie}, I.~B. and {Mosekj{\ae}r}, K.~A. and {Paulsen}, M.~F.~K. and {Sukstorf}, M.~N. and {Wilson}, S.~N. and {{\O}rgaard}, S.~K.~K.},
        title = "{Spectroscopic classification of a complete sample of astrometrically-selected quasar candidates using Gaia DR2}",
      journal = {\aap},
         year = 2020,
        month = dec,
       volume = {644},
          eid = {A17},
        pages = {A17},
          doi = {10.1051/0004-6361/202039262},
archivePrefix = {arXiv},
       eprint = {2010.05934},
 primaryClass = {astro-ph.GA},
       adsurl = {https://ui.adsabs.harvard.edu/abs/2020A&A...644A..17H}
}

@ARTICLE{1986ApJS...61..249W,
       author = {{Wolfe}, A.~M. and {Turnshek}, D.~A. and {Smith}, H.~E. and {Cohen}, R.~D.},
        title = "{Damped Lyman-Alpha Absorption by Disk Galaxies with Large Redshifts. I. The Lick Survey}",
      journal = {\apjs},
         year = 1986,
        month = jun,
       volume = {61},
        pages = {249},
          doi = {10.1086/191114},
       adsurl = {https://ui.adsabs.harvard.edu/abs/1986ApJS...61..249W}
}

@ARTICLE{2012ApJ...753...30S,
       author = {{Stern}, Daniel and {Assef}, Roberto J. and {Benford}, Dominic J. and {Blain}, Andrew and {Cutri}, Roc and {Dey}, Arjun and {Eisenhardt}, Peter and {Griffith}, Roger L. and {Jarrett}, T.~H. and {Lake}, Sean and {Masci}, Frank and {Petty}, Sara and {Stanford}, S.~A. and {Tsai}, Chao-Wei and {Wright}, E.~L. and {Yan}, Lin and {Harrison}, Fiona and {Madsen}, Kristin},
        title = "{Mid-infrared Selection of Active Galactic Nuclei with the Wide-Field Infrared Survey Explorer. I. Characterizing WISE-selected Active Galactic Nuclei in COSMOS}",
      journal = {\apj},
         year = 2012,
        month = jul,
       volume = {753},
       number = {1},
          eid = {30},
        pages = {30},
          doi = {10.1088/0004-637X/753/1/30},
archivePrefix = {arXiv},
       eprint = {1205.0811},
 primaryClass = {astro-ph.CO},
       adsurl = {https://ui.adsabs.harvard.edu/abs/2012ApJ...753...30S}
}

@ARTICLE{2000MNRAS.312..827W,
       author = {{Warren}, S.~J. and {Hewett}, P.~C. and {Foltz}, C.~B.},
        title = "{The KX method for producing K-band flux-limited samples of quasars}",
      journal = {\mnras},
         year = 2000,
        month = mar,
       volume = {312},
       number = {4},
        pages = {827-832},
          doi = {10.1046/j.1365-8711.2000.03206.x},
archivePrefix = {arXiv},
       eprint = {astro-ph/9911064},
 primaryClass = {astro-ph},
       adsurl = {https://ui.adsabs.harvard.edu/abs/2000MNRAS.312..827W}
}

@ARTICLE{1965ApJ...142.1307S,
       author = {{Sandage}, Allan and {V{\'e}ron}, Philippe and {Wyndham}, John D.},
        title = "{Optical Identification of New Quasi-Stellar Radio Sources.}",
      journal = {\apj},
         year = 1965,
        month = oct,
       volume = {142},
        pages = {1307-1311},
          doi = {10.1086/148415},
       adsurl = {https://ui.adsabs.harvard.edu/abs/1965ApJ...142.1307S}
}

@ARTICLE{1990AuJPh..43..227P,
       author = {{Pettini}, Max and {Hunstead}, Richard W.},
        title = "{Metal enrichment, dust and star formation in high-redshift galaxies.}",
      journal = {Australian Journal of Physics},
         year = 1990,
        month = jan,
       volume = {43},
        pages = {227},
          doi = {10.1071/PH900227},
       adsurl = {https://ui.adsabs.harvard.edu/abs/1990AuJPh..43..227P}
}

@ARTICLE{2019A&A...625L...9G,
       author = {{Geier}, S.~J. and {Heintz}, K.~E. and {Fynbo}, J.~P.~U. and {Ledoux}, C. and {Christensen}, L. and {Jakobsson}, P. and {Krogager}, J. -K. and {Milvang-Jensen}, B. and {M{\o}ller}, P. and {Noterdaeme}, P.},
        title = "{Gaia-assisted selection of a quasar reddened by dust in an extremely strong damped Lyman-{\ensuremath{\alpha}} absorber at z = 2.226}",
      journal = {\aap},
         year = 2019,
        month = may,
       volume = {625},
          eid = {L9},
        pages = {L9},
          doi = {10.1051/0004-6361/201935108},
archivePrefix = {arXiv},
       eprint = {1904.01686},
 primaryClass = {astro-ph.GA},
       adsurl = {https://ui.adsabs.harvard.edu/abs/2019A&A...625L...9G}
}

@ARTICLE{2015A&A...578A..91H,
       author = {{Heintz}, K.~E. and {Fynbo}, J.~P.~U. and {H{\o}g}, E.},
        title = "{A study of purely astrometric selection of extragalactic point sources with Gaia}",
      journal = {\aap},
         year = 2015,
        month = jun,
       volume = {578},
          eid = {A91},
        pages = {A91},
          doi = {10.1051/0004-6361/201526038},
archivePrefix = {arXiv},
       eprint = {1503.02874},
 primaryClass = {astro-ph.HE},
       adsurl = {https://ui.adsabs.harvard.edu/abs/2015A&A...578A..91H}
}

@ARTICLE{2016MNRAS.455.2698K,
       author = {{Krogager}, J. -K. and {Fynbo}, J.~P.~U. and {Noterdaeme}, P. and {Zafar}, T. and {M{\o}ller}, P. and {Ledoux}, C. and {Kr{\"u}hler}, T. and {Stockton}, A.},
        title = "{A quasar reddened by a sub-parsec-sized, metal-rich and dusty cloud in a damped Lyman {\ensuremath{\alpha}} absorber at z = 2.13}",
      journal = {\mnras},
         year = 2016,
        month = jan,
       volume = {455},
       number = {3},
        pages = {2698-2711},
          doi = {10.1093/mnras/stv2346},
archivePrefix = {arXiv},
       eprint = {1510.04695},
 primaryClass = {astro-ph.GA},
       adsurl = {https://ui.adsabs.harvard.edu/abs/2016MNRAS.455.2698K}
}

@ARTICLE{2016ApJ...832...49K,
       author = {{Krogager}, J. -K. and {Fynbo}, J.~P.~U. and {Heintz}, K.~E. and {Geier}, S. and {Ledoux}, C. and {M{\o}ller}, P. and {Noterdaeme}, P. and {Venemans}, B.~P. and {Vestergaard}, M.},
        title = "{The Extended High A(V) Quasar Survey: Searching for Dusty Absorbers toward Mid-infrared-selected Quasars}",
      journal = {\apj},
         year = 2016,
        month = nov,
       volume = {832},
       number = {1},
          eid = {49},
        pages = {49},
          doi = {10.3847/0004-637X/832/1/49},
archivePrefix = {arXiv},
       eprint = {1608.08404},
 primaryClass = {astro-ph.GA},
       adsurl = {https://ui.adsabs.harvard.edu/abs/2016ApJ...832...49K}
}

@ARTICLE{1963Natur.197.1040S,
       author = {{Schmidt}, M.},
        title = "{3C 273 : A Star-Like Object with Large Red-Shift}",
      journal = {\nat},
         year = 1963,
        month = mar,
       volume = {197},
       number = {4872},
        pages = {1040},
          doi = {10.1038/1971040a0},
       adsurl = {https://ui.adsabs.harvard.edu/abs/1963Natur.197.1040S}
}

@ARTICLE{1964ApJ...140....1G,
       author = {{Greenstein}, Jesse L. and {Schmidt}, Maarten},
        title = "{The Quasi-Stellar Radio Sources 3C 48 and 3C 273.}",
      journal = {\apj},
         year = 1964,
        month = jul,
       volume = {140},
        pages = {1},
          doi = {10.1086/147889},
       adsurl = {https://ui.adsabs.harvard.edu/abs/1964ApJ...140....1G}
}

@ARTICLE{2013A&A...560A..26Z,
       author = {{Zafar}, Tayyaba and {Watson}, Darach},
        title = "{The metals-to-dust ratio to very low metallicities using GRB and QSO absorbers; extremely rapid dust formation}",
      journal = {\aap},
         year = 2013,
        month = dec,
       volume = {560},
          eid = {A26},
        pages = {A26},
          doi = {10.1051/0004-6361/201321413},
archivePrefix = {arXiv},
       eprint = {1303.1141},
 primaryClass = {astro-ph.CO},
       adsurl = {https://ui.adsabs.harvard.edu/abs/2013A&A...560A..26Z}
}

@ARTICLE{2013MNRAS.430.2680M,
       author = {{M{\o}ller}, P. and {Fynbo}, J.~P.~U. and {Ledoux}, C. and {Nilsson}, K.~K.},
        title = "{Mass-metallicity relation from z = 5 to the present: evidence for a transition in the mode of galaxy growth at z = 2.6 due to the end of sustained primordial gas infall}",
      journal = {\mnras},
         year = 2013,
        month = apr,
       volume = {430},
       number = {4},
        pages = {2680-2687},
          doi = {10.1093/mnras/stt067},
archivePrefix = {arXiv},
       eprint = {1301.5013},
 primaryClass = {astro-ph.CO},
       adsurl = {https://ui.adsabs.harvard.edu/abs/2013MNRAS.430.2680M}
}

@ARTICLE{2006A&A...457...71L,
       author = {{Ledoux}, C. and {Petitjean}, P. and {Fynbo}, J.~P.~U. and {M{\o}ller}, P. and {Srianand}, R.},
        title = "{Velocity-metallicity correlation for high-z DLA galaxies: evidence of a mass-metallicity relation?}",
      journal = {\aap},
         year = 2006,
        month = oct,
       volume = {457},
       number = {1},
        pages = {71-78},
          doi = {10.1051/0004-6361:20054242},
archivePrefix = {arXiv},
       eprint = {astro-ph/0606185},
 primaryClass = {astro-ph},
       adsurl = {https://ui.adsabs.harvard.edu/abs/2006A&A...457...71L}
}

@ARTICLE{2020MNRAS.497.1946B,
       author = {{Balashev}, S.~A. and {Ledoux}, C. and {Noterdaeme}, P. and {Srianand}, R. and {Petitjean}, P. and {Gupta}, N.},
        title = "{Nature of the DLA towards Q 0528-250: High pressure and strong UV field revealed by excitation of C I, H$_{2}$, and Si II}",
      journal = {\mnras},
         year = 2020,
        month = sep,
       volume = {497},
       number = {2},
        pages = {1946-1956},
          doi = {10.1093/mnras/staa2108},
archivePrefix = {arXiv},
       eprint = {2007.07707},
 primaryClass = {astro-ph.GA},
       adsurl = {https://ui.adsabs.harvard.edu/abs/2020MNRAS.497.1946B}
}

@ARTICLE{2018A&A...612A..58N,
       author = {{Noterdaeme}, P. and {Ledoux}, C. and {Zou}, S. and {Petitjean}, P. and {Srianand}, R. and {Balashev}, S. and {L{\'o}pez}, S.},
        title = "{Spotting high-z molecular absorbers using neutral carbon. Results from a complete spectroscopic survey with the VLT}",
      journal = {\aap},
         year = 2018,
        month = apr,
       volume = {612},
          eid = {A58},
        pages = {A58},
          doi = {10.1051/0004-6361/201732266},
archivePrefix = {arXiv},
       eprint = {1801.08357},
 primaryClass = {astro-ph.GA},
       adsurl = {https://ui.adsabs.harvard.edu/abs/2018A&A...612A..58N}
}

@ARTICLE{2015A&A...580A...8L,
       author = {{Ledoux}, C. and {Noterdaeme}, P. and {Petitjean}, P. and {Srianand}, R.},
        title = "{Neutral atomic-carbon quasar absorption-line systems at z> 1.5. Sample selection, H i content, reddening, and 2175 {\r{A}} extinction feature}",
      journal = {\aap},
         year = 2015,
        month = aug,
       volume = {580},
          eid = {A8},
        pages = {A8},
          doi = {10.1051/0004-6361/201424122},
archivePrefix = {arXiv},
       eprint = {1504.07254},
 primaryClass = {astro-ph.GA},
       adsurl = {https://ui.adsabs.harvard.edu/abs/2015A&A...580A...8L}
}

@ARTICLE{2016A&A...585A..87S,
       author = {{Selsing}, J. and {Fynbo}, J.~P.~U. and {Christensen}, L. and {Krogager}, J. -K.},
        title = "{An X-Shooter composite of bright 1 < z < 2 quasars from UV to infrared}",
      journal = {\aap},
         year = 2016,
        month = jan,
       volume = {585},
          eid = {A87},
        pages = {A87},
          doi = {10.1051/0004-6361/201527096},
archivePrefix = {arXiv},
       eprint = {1510.08058},
 primaryClass = {astro-ph.GA},
       adsurl = {https://ui.adsabs.harvard.edu/abs/2016A&A...585A..87S}
}

@ARTICLE{2007ApJ...663..320F,
       author = {{Fitzpatrick}, E.~L. and {Massa}, D.},
        title = "{An Analysis of the Shapes of Interstellar Extinction Curves. V. The IR-through-UV Curve Morphology}",
      journal = {\apj},
         year = 2007,
        month = jul,
       volume = {663},
       number = {1},
        pages = {320-341},
          doi = {10.1086/518158},
archivePrefix = {arXiv},
       eprint = {0705.0154},
 primaryClass = {astro-ph},
       adsurl = {https://ui.adsabs.harvard.edu/abs/2007ApJ...663..320F}
}

@article{Tepper_Garc_a_2006,
   title={Voigt profile fitting to quasar absorption lines: an analytic approximation to the Voigt-Hjerting function: A new method to compute Voigt profiles},
   volume={369},
   ISSN={0035-8711},
   url={http://dx.doi.org/10.1111/j.1365-2966.2006.10450.x},
   DOI={10.1111/j.1365-2966.2006.10450.x},
   number={4},
   journal={Monthly Notices of the Royal Astronomical Society},
   publisher={Oxford University Press (OUP)},
   author={Tepper García, Thorsten},
   year={2006},
   month=jun, pages={2025–2035} }

@ARTICLE{2003ApJ...594..279G,
       author = {{Gordon}, Karl D. and {Clayton}, Geoffrey C. and {Misselt}, K.~A. and {Landolt}, Arlo U. and {Wolff}, Michael J.},
        title = "{A Quantitative Comparison of the Small Magellanic Cloud, Large Magellanic Cloud, and Milky Way Ultraviolet to Near-Infrared Extinction Curves}",
      journal = {\apj},
         year = 2003,
        month = sep,
       volume = {594},
       number = {1},
        pages = {279-293},
          doi = {10.1086/376774},
archivePrefix = {arXiv},
       eprint = {astro-ph/0305257},
 primaryClass = {astro-ph},
       adsurl = {https://ui.adsabs.harvard.edu/abs/2003ApJ...594..279G}
}

@ARTICLE{2018arXiv180301187K,
       author = {{Krogager}, Jens-Kristian},
        title = "{VoigtFit: A Python package for Voigt profile fitting}",
      journal = {arXiv e-prints},
         year = 2018,
        month = mar,
          eid = {arXiv:1803.01187},
        pages = {arXiv:1803.01187},
          doi = {10.48550/arXiv.1803.01187},
archivePrefix = {arXiv},
       eprint = {1803.01187},
 primaryClass = {astro-ph.IM},
       adsurl = {https://ui.adsabs.harvard.edu/abs/2018arXiv180301187K}
}

@ARTICLE{1990ApJ...363...57M,
       author = {{Meyer}, David M. and {Roth}, Katherine C.},
        title = "{Observations of Nickel, Chromium, and Zinc in QSO Absorption-Line Systems}",
      journal = {\apj},
         year = 1990,
        month = nov,
       volume = {363},
        pages = {57},
          doi = {10.1086/169319},
       adsurl = {https://ui.adsabs.harvard.edu/abs/1990ApJ...363...57M}
}

@ARTICLE{1994ApJ...426...79P,
       author = {{Pettini}, Max and {Smith}, Linda J. and {Hunstead}, Richard W. and {King}, David L.},
        title = "{Metal Enrichment, Dust, and Star Formation in Galaxies at High Redshifts. III. Zn and CR Abundances for 17 Damped Lyman-Alpha Systems}",
      journal = {\apj},
         year = 1994,
        month = may,
       volume = {426},
        pages = {79},
          doi = {10.1086/174041},
       adsurl = {https://ui.adsabs.harvard.edu/abs/1994ApJ...426...79P}
}

@ARTICLE{2016A&A...596A..97D,
       author = {{De Cia}, A. and {Ledoux}, C. and {Mattsson}, L. and {Petitjean}, P. and {Srianand}, R. and {Gavignaud}, I. and {Jenkins}, E.~B.},
        title = "{Dust-depletion sequences in damped Lyman-{\ensuremath{\alpha}} absorbers. A unified picture from low-metallicity systems to the Galaxy}",
      journal = {\aap},
         year = 2016,
        month = dec,
       volume = {596},
          eid = {A97},
        pages = {A97},
          doi = {10.1051/0004-6361/201527895},
archivePrefix = {arXiv},
       eprint = {1608.08621},
 primaryClass = {astro-ph.GA},
       adsurl = {https://ui.adsabs.harvard.edu/abs/2016A&A...596A..97D}
}

@ARTICLE{2010MNRAS.408.2128F,
       author = {{Fynbo}, J.~P.~U. and {Laursen}, P. and {Ledoux}, C. and {M{\o}ller}, P. and {Durgapal}, A.~K. and {Goldoni}, P. and {Gullberg}, B. and {Kaper}, L. and {Maund}, J. and {Noterdaeme}, P. and {{\"O}stlin}, G. and {Strandet}, M.~L. and {Toft}, S. and {Vreeswijk}, P.~M. and {Zafar}, T.},
        title = "{Galaxy counterparts of metal-rich damped Ly{\ensuremath{\alpha}} absorbers - I. The case of the z = 2.35 DLA towards Q2222-0946}",
      journal = {\mnras},
         year = 2010,
        month = nov,
       volume = {408},
       number = {4},
        pages = {2128-2136},
          doi = {10.1111/j.1365-2966.2010.17294.x},
archivePrefix = {arXiv},
       eprint = {1002.4626},
 primaryClass = {astro-ph.CO},
       adsurl = {https://ui.adsabs.harvard.edu/abs/2010MNRAS.408.2128F}
}

@ARTICLE{2019A&A...623A..92S,
       author = {{Selsing}, J. and {Malesani}, D. and {Goldoni}, P. and {Fynbo}, J.~P.~U. and {Kr{\"u}hler}, T. and {Antonelli}, L.~A. and {Arabsalmani}, M. and {Bolmer}, J. and {Cano}, Z. and {Christensen}, L. and {Covino}, S. and {D'Avanzo}, P. and {D'Elia}, V. and {De Cia}, A. and {de Ugarte Postigo}, A. and {Flores}, H. and {Friis}, M. and {Gomboc}, A. and {Greiner}, J. and {Groot}, P. and {Hammer}, F. and {Hartoog}, O.~E. and {Heintz}, K.~E. and {Hjorth}, J. and {Jakobsson}, P. and {Japelj}, J. and {Kann}, D.~A. and {Kaper}, L. and {Ledoux}, C. and {Leloudas}, G. and {Levan}, A.~J. and {Maiorano}, E. and {Melandri}, A. and {Milvang-Jensen}, B. and {Palazzi}, E. and {Palmerio}, J.~T. and {Perley}, D.~A. and {Pian}, E. and {Piranomonte}, S. and {Pugliese}, G. and {S{\'a}nchez-Ram{\'\i}rez}, R. and {Savaglio}, S. and {Schady}, P. and {Schulze}, S. and {Sollerman}, J. and {Sparre}, M. and {Tagliaferri}, G. and {Tanvir}, N.~R. and {Th{\"o}ne}, C.~C. and {Vergani}, S.~D. and {Vreeswijk}, P. and {Watson}, D. and {Wiersema}, K. and {Wijers}, R. and {Xu}, D. and {Zafar}, T.},
        title = "{The X-shooter GRB afterglow legacy sample (XS-GRB)}",
      journal = {\aap},
         year = 2019,
        month = mar,
       volume = {623},
          eid = {A92},
        pages = {A92},
          doi = {10.1051/0004-6361/201832835},
archivePrefix = {arXiv},
       eprint = {1802.07727},
 primaryClass = {astro-ph.HE},
       adsurl = {https://ui.adsabs.harvard.edu/abs/2019A&A...623A..92S}
}

@ARTICLE{2003ApJ...593..215W,
       author = {{Wolfe}, Arthur M. and {Prochaska}, Jason X. and {Gawiser}, Eric},
        title = "{C II* Absorption in Damped Ly{\ensuremath{\alpha}} Systems. I. Star Formation Rates in a Two-Phase Medium}",
      journal = {\apj},
         year = 2003,
        month = aug,
       volume = {593},
       number = {1},
        pages = {215-234},
          doi = {10.1086/376520},
archivePrefix = {arXiv},
       eprint = {astro-ph/0304040},
 primaryClass = {astro-ph},
       adsurl = {https://ui.adsabs.harvard.edu/abs/2003ApJ...593..215W}
}

@ARTICLE{2003ApJ...593..235W,
       author = {{Wolfe}, Arthur M. and {Gawiser}, Eric and {Prochaska}, Jason X.},
        title = "{C II* Absorption in Damped Ly{\ensuremath{\alpha}} Systems. II. A New Window on the Star Formation History of the Universe}",
      journal = {\apj},
         year = 2003,
        month = aug,
       volume = {593},
       number = {1},
        pages = {235-257},
          doi = {10.1086/376521},
archivePrefix = {arXiv},
       eprint = {astro-ph/0304042},
 primaryClass = {astro-ph},
       adsurl = {https://ui.adsabs.harvard.edu/abs/2003ApJ...593..235W}
}

@ARTICLE{1981ApJ...246L.147L,
       author = {{Liszt}, H.~S.},
        title = "{The carbon abundance in diffuse interstellar clouds.}",
      journal = {\apjl},
         year = 1981,
        month = jun,
       volume = {246},
        pages = {L147-L150},
          doi = {10.1086/183572},
       adsurl = {https://ui.adsabs.harvard.edu/abs/1981ApJ...246L.147L}
}

@INPROCEEDINGS{2010SPIE.7737E..28M,
       author = {{Modigliani}, Andrea and {Goldoni}, Paolo and {Royer}, Fr{\'e}d{\'e}ric and {Haigron}, Regis and {Guglielmi}, Laurent and {Fran{\c{c}}ois}, Patrick and {Horrobin}, Matthew and {Bristow}, Paul and {Vernet}, Joel and {Moehler}, Sabine and {Kerber}, Florian and {Ballester}, Pascal and {Mason}, Elena and {Christensen}, Lise},
        title = "{The X-shooter pipeline}",
    booktitle = {Observatory Operations: Strategies, Processes, and Systems III},
         year = 2010,
       editor = {{Silva}, David R. and {Peck}, Alison B. and {Soifer}, B. Thomas},
       series = {Society of Photo-Optical Instrumentation Engineers (SPIE) Conference Series},
       volume = {7737},
        month = jul,
          eid = {773728},
        pages = {773728},
          doi = {10.1117/12.857211},
       adsurl = {https://ui.adsabs.harvard.edu/abs/2010SPIE.7737E..28M}
}

@ARTICLE{2020ApJ...893...14D,
       author = {{Dix}, Cooper and {Shemmer}, Ohad and {Brotherton}, Michael S. and {Green}, Richard F. and {Mason}, Michelle and {Myers}, Adam D.},
        title = "{Prescriptions for Correcting Ultraviolet-based Redshifts for Luminous Quasars at High Redshift}",
      journal = {\apj},
         year = 2020,
        month = apr,
       volume = {893},
       number = {1},
          eid = {14},
        pages = {14},
          doi = {10.3847/1538-4357/ab77b6},
archivePrefix = {arXiv},
       eprint = {2002.08472},
 primaryClass = {astro-ph.GA},
       adsurl = {https://ui.adsabs.harvard.edu/abs/2020ApJ...893...14D}
}

@ARTICLE{2016ApJ...831....7S,
       author = {{Shen}, Yue and {Brandt}, W.~N. and {Richards}, Gordon T. and {Denney}, Kelly D. and {Greene}, Jenny E. and {Grier}, C.~J. and {Ho}, Luis C. and {Peterson}, Bradley M. and {Petitjean}, Patrick and {Schneider}, Donald P. and {Tao}, Charling and {Trump}, Jonathan R.},
        title = "{The Sloan Digital Sky Survey Reverberation Mapping Project: Velocity Shifts of Quasar Emission Lines}",
      journal = {\apj},
         year = 2016,
        month = nov,
       volume = {831},
       number = {1},
          eid = {7},
        pages = {7},
          doi = {10.3847/0004-637X/831/1/7},
archivePrefix = {arXiv},
       eprint = {1602.03894},
 primaryClass = {astro-ph.GA},
       adsurl = {https://ui.adsabs.harvard.edu/abs/2016ApJ...831....7S}
}

@ARTICLE{1992ApJS...79....1T,
       author = {{Tytler}, David and {Fan}, Xiao-Ming},
        title = "{Systematic QSO Emission-Line Velocity Shifts and New Unbiased Redshifts}",
      journal = {\apjs},
         year = 1992,
        month = mar,
       volume = {79},
        pages = {1},
          doi = {10.1086/191642},
       adsurl = {https://ui.adsabs.harvard.edu/abs/1992ApJS...79....1T}
}

@ARTICLE{2010MNRAS.405.2302H,
       author = {{Hewett}, Paul C. and {Wild}, Vivienne},
        title = "{Improved redshifts for SDSS quasar spectra}",
      journal = {\mnras},
         year = 2010,
        month = jul,
       volume = {405},
       number = {4},
        pages = {2302-2316},
          doi = {10.1111/j.1365-2966.2010.16648.x},
archivePrefix = {arXiv},
       eprint = {1003.3017},
 primaryClass = {astro-ph.CO},
       adsurl = {https://ui.adsabs.harvard.edu/abs/2010MNRAS.405.2302H}
}

@ARTICLE{1978PhyS...17..217W,
       author = {{Weymann}, R.~J. and {Williams}, R.~E.},
        title = "{A summary of the properties of Z$_{abs}$ > Z$_{em}$ systems and some comments on the origins of QSO absorption line systems.}",
      journal = {\physscr},
         year = 1978,
        month = mar,
       volume = {17},
        pages = {217-223},
          doi = {10.1088/0031-8949/17/3/014},
       adsurl = {https://ui.adsabs.harvard.edu/abs/1978PhyS...17..217W}
}

@ARTICLE{ Noterdaeme2017,
 author = {{Noterdaeme}, P. and {Krogager}, J.-K. and {Balashev}, S. and 
{Ge}, J. and {Gupta}, N. and {Kr{\"u}hler}, T. and {Ledoux}, C. and 
{Murphy}, M.~T. and {P{\^a}ris}, I. and {Petitjean}, P. and 
{Rahmani}, H. and {Srianand}, R. and {Ubachs}, W.},
 title = "{Discovery of a Perseus-like cloud in the early Universe. H I-to-H$_{2}$ transition, carbon monoxide and small dust grains at \ensuremath{z_{abs} = 2.53} towards the quasar J0000+0048}",
 journal = {\aap},
archivePrefix = "arXiv",
 eprint = {1609.01422},
 year = 2017,
 month = jan,
 volume = 597,
 eid = {A82},
 pages = {A82},
 doi = {10.1051/0004-6361/201629173},
 adsurl = {http://adsabs.harvard.edu/abs/2017A%26A...597A..82N}
}

@ARTICLE{2011MNRAS.418L..74D,
       author = {{de La Rosa}, Ignacio G. and {La Barbera}, Francesco and {Ferreras}, Ignacio and {de Carvalho}, Reinaldo R.},
        title = "{The link between the star formation history and [{\ensuremath{\alpha}}/Fe ]}",
      journal = {\mnras},
         year = 2011,
        month = nov,
       volume = {418},
       number = {1},
        pages = {L74-L78},
          doi = {10.1111/j.1745-3933.2011.01146.x},
archivePrefix = {arXiv},
       eprint = {1109.0005},
 primaryClass = {astro-ph.CO},
       adsurl = {https://ui.adsabs.harvard.edu/abs/2011MNRAS.418L..74D}
}

@ARTICLE{2005MNRAS.362..549S,
       author = {{Srianand}, Raghunathan and {Petitjean}, Patrick and {Ledoux}, C{\'e}dric and {Ferland}, Gary and {Shaw}, Gargi},
        title = "{The VLT-UVES survey for molecular hydrogen in high-redshift damped Lyman {\ensuremath{\alpha}} systems: physical conditions in the neutral gas}",
      journal = {\mnras},
         year = 2005,
        month = sep,
       volume = {362},
       number = {2},
        pages = {549-568},
          doi = {10.1111/j.1365-2966.2005.09324.x},
archivePrefix = {arXiv},
       eprint = {astro-ph/0506555},
 primaryClass = {astro-ph},
       adsurl = {https://ui.adsabs.harvard.edu/abs/2005MNRAS.362..549S}
}

@ARTICLE{2006MNRAS.365L...1R,
       author = {{Roy}, Nirupam and {Chengalur}, Jayaram N. and {Srianand}, Raghunathan},
        title = "{A multiwavelength investigation of the temperature of the cold neutral medium}",
      journal = {\mnras},
         year = 2006,
        month = jan,
       volume = {365},
       number = {1},
        pages = {L1-L5},
          doi = {10.1111/j.1745-3933.2005.00114.x},
archivePrefix = {arXiv},
       eprint = {astro-ph/0511297},
 primaryClass = {astro-ph},
       adsurl = {https://ui.adsabs.harvard.edu/abs/2006MNRAS.365L...1R}
}

@ARTICLE{1977ApJ...216..291S,
       author = {{Savage}, B.~D. and {Bohlin}, R.~C. and {Drake}, J.~F. and {Budich}, W.},
        title = "{A survey of interstellar molecular hydrogen. I.}",
      journal = {\apj},
         year = 1977,
        month = aug,
       volume = {216},
        pages = {291-307},
          doi = {10.1086/155471},
       adsurl = {https://ui.adsabs.harvard.edu/abs/1977ApJ...216..291S}
}

@ARTICLE{2015MNRAS.448.2840M,
       author = {{Muzahid}, S. and {Srianand}, R. and {Charlton}, J.},
        title = "{An HST/COS survey of molecular hydrogen in DLAs \& sub-DLAs at z < 1: molecular fraction and excitation temperature}",
      journal = {\mnras},
         year = 2015,
        month = apr,
       volume = {448},
       number = {3},
        pages = {2840-2853},
          doi = {10.1093/mnras/stv133},
archivePrefix = {arXiv},
       eprint = {1410.3828},
 primaryClass = {astro-ph.GA},
       adsurl = {https://ui.adsabs.harvard.edu/abs/2015MNRAS.448.2840M}
}

@INPROCEEDINGS{2002AAS...201.7702R,
       author = {{Rachford}, B.~L. and {Baker}, E.~J. and {Snow}, T.~P.},
        title = "{FUSE Observations of Rotationally Excited H$_{2}$ in Translucent Lines of Sight}",
    booktitle = {American Astronomical Society Meeting Abstracts},
         year = 2002,
       series = {American Astronomical Society Meeting Abstracts},
       volume = {201},
        month = dec,
          eid = {77.02},
        pages = {77.02},
       adsurl = {https://ui.adsabs.harvard.edu/abs/2002AAS...201.7702R}
}

@ARTICLE{2012ApJ...745..173W,
       author = {{Welty}, Daniel E. and {Xue}, Rui and {Wong}, Tony},
        title = "{Interstellar H I and H$_{2}$ in the Magellanic Clouds: An Expanded Sample Based on Ultraviolet Absorption-line Data}",
      journal = {\apj},
         year = 2012,
        month = feb,
       volume = {745},
       number = {2},
          eid = {173},
        pages = {173},
          doi = {10.1088/0004-637X/745/2/173},
archivePrefix = {arXiv},
       eprint = {1111.3674},
 primaryClass = {astro-ph.GA},
       adsurl = {https://ui.adsabs.harvard.edu/abs/2012ApJ...745..173W}
}

@ARTICLE{2010ApJ...722..460J,
       author = {{Jorgenson}, Regina A. and {Wolfe}, Arthur M. and {Prochaska}, J. Xavier},
        title = "{Understanding Physical Conditions in High-redshift Galaxies Through C I Fine Structure Lines: Data and Methodology}",
      journal = {\apj},
         year = 2010,
        month = oct,
       volume = {722},
       number = {1},
        pages = {460-490},
          doi = {10.1088/0004-637X/722/1/460},
archivePrefix = {arXiv},
       eprint = {1008.4676},
 primaryClass = {astro-ph.CO},
       adsurl = {https://ui.adsabs.harvard.edu/abs/2010ApJ...722..460J}
}

@ARTICLE{2018A&A...615A..43H,
       author = {{Heintz}, K.~E. and {Fynbo}, J.~P.~U. and {Ledoux}, C. and {Jakobsson}, P. and {M{\o}ller}, P. and {Christensen}, L. and {Geier}, S. and {Krogager}, J. -K. and {Noterdaeme}, P.},
        title = "{A quasar hiding behind two dusty absorbers. Quantifying the selection bias of metal-rich, damped Ly{\ensuremath{\alpha}} absorption systems}",
      journal = {\aap},
         year = 2018,
        month = jul,
       volume = {615},
          eid = {A43},
        pages = {A43},
          doi = {10.1051/0004-6361/201731964},
archivePrefix = {arXiv},
       eprint = {1803.09805},
 primaryClass = {astro-ph.GA},
       adsurl = {https://ui.adsabs.harvard.edu/abs/2018A&A...615A..43H}
}

@ARTICLE{2018A&A...615L...8H,
       author = {{Heintz}, K.~E. and {Fynbo}, J.~P.~U. and {H{\o}g}, E. and {M{\o}ller}, P. and {Krogager}, J. -K. and {Geier}, S. and {Jakobsson}, P. and {Christensen}, L.},
        title = "{Unidentified quasars among stationary objects from Gaia DR2}",
      journal = {\aap},
         year = 2018,
        month = jul,
       volume = {615},
          eid = {L8},
        pages = {L8},
          doi = {10.1051/0004-6361/201833396},
archivePrefix = {arXiv},
       eprint = {1805.03394},
 primaryClass = {astro-ph.GA},
       adsurl = {https://ui.adsabs.harvard.edu/abs/2018A&A...615L...8H}
}

@ARTICLE{2023A&A...679A..91H,
       author = {{Heintz}, K.~E. and {De Cia}, A. and {Th{\"o}ne}, C.~C. and {Krogager}, J. -K. and {Yates}, R.~M. and {Vejlgaard}, S. and {Konstantopoulou}, C. and {Fynbo}, J.~P.~U. and {Watson}, D. and {Narayanan}, D. and {Wilson}, S.~N. and {Arabsalmani}, M. and {Campana}, S. and {D'Elia}, V. and {De Pasquale}, M. and {Hartmann}, D.~H. and {Izzo}, L. and {Jakobsson}, P. and {Kouveliotou}, C. and {Levan}, A. and {Li}, Q. and {Malesani}, D.~B. and {Melandri}, A. and {Milvang-Jensen}, B. and {M{\o}ller}, P. and {Palazzi}, E. and {Palmerio}, J. and {Petitjean}, P. and {Pugliese}, G. and {Rossi}, A. and {Saccardi}, A. and {Salvaterra}, R. and {Savaglio}, S. and {Schady}, P. and {Stratta}, G. and {Tanvir}, N.~R. and {de Ugarte Postigo}, A. and {Vergani}, S.~D. and {Wiersema}, K. and {Wijers}, R.~A.~M.~J. and {Zafar}, T.},
        title = "{The cosmic buildup of dust and metals. Accurate abundances from GRB-selected star-forming galaxies at 1.7 < z < 6.3}",
      journal = {\aap},
         year = 2023,
        month = nov,
       volume = {679},
          eid = {A91},
        pages = {A91},
          doi = {10.1051/0004-6361/202347418},
archivePrefix = {arXiv},
       eprint = {2308.14812},
 primaryClass = {astro-ph.GA},
       adsurl = {https://ui.adsabs.harvard.edu/abs/2023A&A...679A..91H}
}

@ARTICLE{2023A&A...673A..89N,
       author = {{Noterdaeme}, P. and {Balashev}, S. and {Cuellar}, R. and {Krogager}, J. -K. and {Combes}, F. and {De Cia}, A. and {Gupta}, N. and {Ledoux}, C. and {L{\'o}pez}, S. and {Srianand}, R.},
        title = "{Proximate molecular quasar absorbers. Chemical enrichment and kinematics of the neutral gas}",
      journal = {\aap},
         year = 2023,
        month = may,
       volume = {673},
          eid = {A89},
        pages = {A89},
          doi = {10.1051/0004-6361/202245554},
archivePrefix = {arXiv},
       eprint = {2302.13108},
 primaryClass = {astro-ph.GA},
       adsurl = {https://ui.adsabs.harvard.edu/abs/2023A&A...673A..89N}
}

@ARTICLE{2025Natur.641.1137B,
       author = {{Balashev}, Sergei and {Noterdaeme}, Pasquier and {Gupta}, Neeraj and {Krogager}, Jens-Kristian and {Combes}, Fran{\c{c}}oise and {L{\'o}pez}, Sebasti{\'a}n and {Petitjean}, Patrick and {Omont}, Alain and {Srianand}, Raghunathan and {Cuellar}, Rodrigo},
        title = "{Quasar radiation transforms the gas in a merging companion galaxy}",
      journal = {\nat},
         year = 2025,
        month = may,
       volume = {641},
       number = {8065},
        pages = {1137-1141},
          doi = {10.1038/s41586-025-08966-4},
archivePrefix = {arXiv},
       eprint = {2505.15766},
 primaryClass = {astro-ph.GA},
       adsurl = {https://ui.adsabs.harvard.edu/abs/2025Natur.641.1137B}
}

@ARTICLE{2024A&A...691A.129K,
       author = {{Konstantopoulou}, Christina and {De Cia}, Annalisa and {Krogager}, Jens-Kristian and {Ledoux}, C{\'e}dric and {Roman-Duval}, Julia and {Jenkins}, Edward B. and {Ramburuth-Hurt}, Tanita and {Velichko}, Anna},
        title = "{DUNE: Dust depletion UNified method across cosmic time and Environments}",
      journal = {\aap},
         year = 2024,
        month = nov,
       volume = {691},
          eid = {A129},
        pages = {A129},
          doi = {10.1051/0004-6361/202451488},
archivePrefix = {arXiv},
       eprint = {2410.06155},
 primaryClass = {astro-ph.GA},
       adsurl = {https://ui.adsabs.harvard.edu/abs/2024A&A...691A.129K}
}

@ARTICLE{2019A&A...621A..20H,
       author = {{Heintz}, K.~E. and {Ledoux}, C. and {Fynbo}, J.~P.~U. and {Jakobsson}, P. and {Noterdaeme}, P. and {Krogager}, J. -K. and {Bolmer}, J. and {M{\o}ller}, P. and {Vergani}, S.~D. and {Watson}, D. and {Zafar}, T. and {De Cia}, A. and {Tanvir}, N.~R. and {Malesani}, D.~B. and {Japelj}, J. and {Covino}, S. and {Kaper}, L.},
        title = "{Cold gas in the early Universe. Survey for neutral atomic-carbon in GRB host galaxies at 1 < z< 6 from optical afterglow spectroscopy}",
      journal = {\aap},
         year = 2019,
        month = jan,
       volume = {621},
          eid = {A20},
        pages = {A20},
          doi = {10.1051/0004-6361/201834246},
archivePrefix = {arXiv},
       eprint = {1810.11064},
 primaryClass = {astro-ph.GA},
       adsurl = {https://ui.adsabs.harvard.edu/abs/2019A&A...621A..20H}
}

@ARTICLE{2019A&A...629A.131H,
       author = {{Heintz}, K.~E. and {Bolmer}, J. and {Ledoux}, C. and {Noterdaeme}, P. and {Krogager}, J. -K. and {Fynbo}, J.~P.~U. and {Jakobsson}, P. and {Covino}, S. and {D'Elia}, V. and {De Pasquale}, M. and {Hartmann}, D.~H. and {Izzo}, L. and {Japelj}, J. and {Kann}, D.~A. and {Kaper}, L. and {Petitjean}, P. and {Rossi}, A. and {Salvaterra}, R. and {Schady}, P. and {Selsing}, J. and {Starling}, R. and {Tanvir}, N.~R. and {Th{\"o}ne}, C.~C. and {de Ugarte Postigo}, A. and {Vergani}, S.~D. and {Watson}, D. and {Wiersema}, K. and {Zafar}, T.},
        title = "{New constraints on the physical conditions in H$_{2}$-bearing GRB-host damped Lyman-{\ensuremath{\alpha}} absorbers}",
      journal = {\aap},
         year = 2019,
        month = sep,
       volume = {629},
          eid = {A131},
        pages = {A131},
          doi = {10.1051/0004-6361/201936250},
archivePrefix = {arXiv},
       eprint = {1908.02309},
 primaryClass = {astro-ph.GA},
       adsurl = {https://ui.adsabs.harvard.edu/abs/2019A&A...629A.131H}
}

@ARTICLE{2020ApJ...889L...7H,
       author = {{Heintz}, Kasper E. and {Watson}, Darach},
        title = "{Direct Measurement of the [C I] Luminosity to Molecular Gas Mass Conversion Factor in High-redshift Star-forming Galaxies}",
      journal = {\apjl},
         year = 2020,
        month = jan,
       volume = {889},
       number = {1},
          eid = {L7},
        pages = {L7},
          doi = {10.3847/2041-8213/ab6733},
archivePrefix = {arXiv},
       eprint = {2001.05770},
 primaryClass = {astro-ph.GA},
       adsurl = {https://ui.adsabs.harvard.edu/abs/2020ApJ...889L...7H}
}

@ARTICLE{1989ApJ...337....7F,
       author = {{Fall}, S. Michael and {Pei}, Yichuan C.},
        title = "{Limits on Dust in Damped Lyman-Alpha Systems and the Obscuration of Quasars}",
      journal = {\apj},
         year = 1989,
        month = feb,
       volume = {337},
        pages = {7},
          doi = {10.1086/167083},
       adsurl = {https://ui.adsabs.harvard.edu/abs/1989ApJ...337....7F}
}

@ARTICLE{1991ApJ...378....6P,
       author = {{Pei}, Yichuan C. and {Fall}, S.~M. and {Bechtold}, Jill},
        title = "{Confirmation of Dust in Damped Lyman-Alpha Systems}",
      journal = {\apj},
         year = 1991,
        month = sep,
       volume = {378},
        pages = {6},
          doi = {10.1086/170401},
       adsurl = {https://ui.adsabs.harvard.edu/abs/1991ApJ...378....6P}
}

@ARTICLE{2009MNRAS.393..557P,
       author = {{Pontzen}, Andrew and {Pettini}, Max},
        title = "{Dust biasing of damped Lyman alpha systems: a Bayesian analysis}",
      journal = {\mnras},
         year = 2009,
        month = feb,
       volume = {393},
       number = {2},
        pages = {557-568},
          doi = {10.1111/j.1365-2966.2008.14193.x},
archivePrefix = {arXiv},
       eprint = {0810.3236},
 primaryClass = {astro-ph},
       adsurl = {https://ui.adsabs.harvard.edu/abs/2009MNRAS.393..557P}
}

\end{document}